# Adaptive control for trailing arm suspension with hydraulic cylinder spring under practical constraints

Seungjun Jung and Hyeongcheol Lee

*Abstract*— This paper presents an adaptive control strategy with LQ control of a quarter-car, which is ameliorated to smoothly follow the dynamically changing reference arm angle, in order to omit the heave sensor in this study. Linearized plant parameters are initially derived from model equation and then estimated by online estimator. LQ control method provides optimal control effort with corresponding estimated plant parameters. The results of the simulation are depicted in time domain. Weighted RMS values of the consequences are suggested to assess the efficiency of the controller. Different road conditions are considered in order to reveal the performance of the controller in detail.

*Index Terms*— Automated Manual Transmission, Synchronizer Gear Shifting, Motor Position Control, Modified Linear Quadratic Tracking, Luenberger Observer

## I. Introduction

Suspension system in any vehicle is important because they isolate a car body from road disturbances in order to provide good ride quality, and enables the wheels to maintain contact with the road, assuring stability and control of the vehicle. Suspension system has three crucial elements such as flexibility (provided by spring), damping, and location of the wheel or axle. Control targets in a suspension system usually fall into two groups: the springing function (i.e., the main springs, dampers, and anti-roll bars), and locating the wheel assemblies and controlling the geometries of their movements under dynamic laden conditions [1]. A suspension with a solid connection between the left and right wheels is called dependent suspension. A suspension is called independent suspension when they let a wheel to move up and down without affecting the opposite wheel. An independent suspension usually categorized into one of four suspension types (trailing arm suspension [2], double A arm suspension [3], McPherson suspension [4], swing arm suspension [5]) regarding their structural scheme [6]. Sometimes, additional stabilizer, antiroll bar, is attached to a solid axle on a vehicle with coil springs [7]. Vehicle models that are used to derive the model equation of a target are quarter-car, half-car, and full-car model [8]. A quarter-car model, the simplest one between three vehicle models, represents the automotive suspension system at each wheel (the motion of the axle and of the vehicle body at each wheel). The next simple vehicle model is half-car, which also concerns pitch motion of the overall vehicle. A full-car model is the most detailed representation of a vehicle model while it is sometimes too complicated to derive. Any model equation that derived using aforementioned vehicle model is used to control the vehicle's suspension system. A suspension system can be sorted to one of three categories: passive semi-active, and active. Passive suspension is composed of springs and dampers placed between the vehicle body and wheel-axle assembly. Semi-active suspension utilizes a variable damper or other variable dissipation component in the automotive system [9, 10, 11]. Active suspension is one in which dampers and metal springs are replaced, or added, with actuators that under computer control, adjust in length in response to road inputs and can maintain zero roll and pitch changes. For optimum performance, such a system needs road preview that is a system that "sees" approaching bumps and anticipates them so it can be proactive rather than reactive [12, 13].

Previous researches on active suspension control significantly enhanced ride quality with using such as fuzzy control [14], preview control [15]. Especially, tremendous enhancement in ride quality is achieved on trailing arm suspension with preview control [15]. However, precedent study mainly focused on fostering ride quality, but some physical constraints, such as maximum range in trailing arm angle and inaccuracy in heave displacement. Also, heave sensors are too costly. In this paper, the main focus is on improving ride quality under restriction in range in arm angle and torque output without using heave sensor. A quarter-car model is used as the plant model and SISO adaptive LQ controller is designed to improve tracking performance for the dynamically changing reference. The target vehicle is similar to the panzers used in previous researches, but not exactly same in that it has hydraulic cylinder spring attached in each of its trailing arm suspension. Also, unlike previous research with panzer greatly reaped benefit in ride quality using wheelbase preview control [15], the target vehicle will less likely to get same benefit since the gap between the front wheel and middle wheel, where the active suspension controller is attached, is significantly larger. This major difference will yield that ride quality is dominantly determined by the front wheel suspension. Thus, wheelbase preview control would not be beneficial to this vehicle. Moreover, as discussed in [15], using look-ahead sensors for preview control will also have some deficiencies. For example, first, they will recognize a heap of leaves as a serious obstacle, while a ditch filled with water might not be detected at all. Second, additional implementation of preview sensors may produce a considerable cost increase. Finally, motion of the vehicle may distort the signals of look-ahead sensors that are attached to the vehicle's body. Thus, heave sensor is omitted and, consequently, different approach is required from the previous research.

Armored vehicle systems, as like high mobility tracked vehicle, require good off-road mobility for high shooting accuracy and survivability in combat situations [16]. As the driving velocity of armored vehicle systems gets larger, any undesirable vibration induced by rough terrain also increases, resulting discomforts on crew members and may have adverse effects on the many delicate tools inside the vehicle. What's more, since vibration in a gun barrel would impedes shooting accuracy, if excessive vibration from the road is not attenuated, this vibration would restricts the maximum vehicle speed and, as a consequence, may reduce not only the survivability but also the operational efficiency in combat situations.

This paper is organized in following sense: Section 2 – modeling of the armored vehicle system, Section 3 – adaptive control scheme and overall control synthesis, Section 4 – a performance comparison between passive and active system and ride quality analysis, and conclusions will be drawn in Section 5. Numerous simulations are given on various environments to verify the performance of the suspension using a commercial multibody dynamics simulation [17] tool, RecurDyn™.

## II. VEHICLE MODEL

### A. Target system

In this study, the target system, a military vehicle, is modeled using RecurDyn™. Figure 1 shows the target system. The vehicle model has totally 6 wheels, 4 wheels' suspension, trailing arm suspension, are actively controlled while 2 wheels, rear wheels, are under passive suspension, double A arm suspension, using damper and spring. Figure 2 and Figure 3 display our suspension system model. Existence of the ordinary passive suspension using damper and spring, as like in last wheel, is crucial, since it is hard to attenuate fast oscillation with only active suspensions.

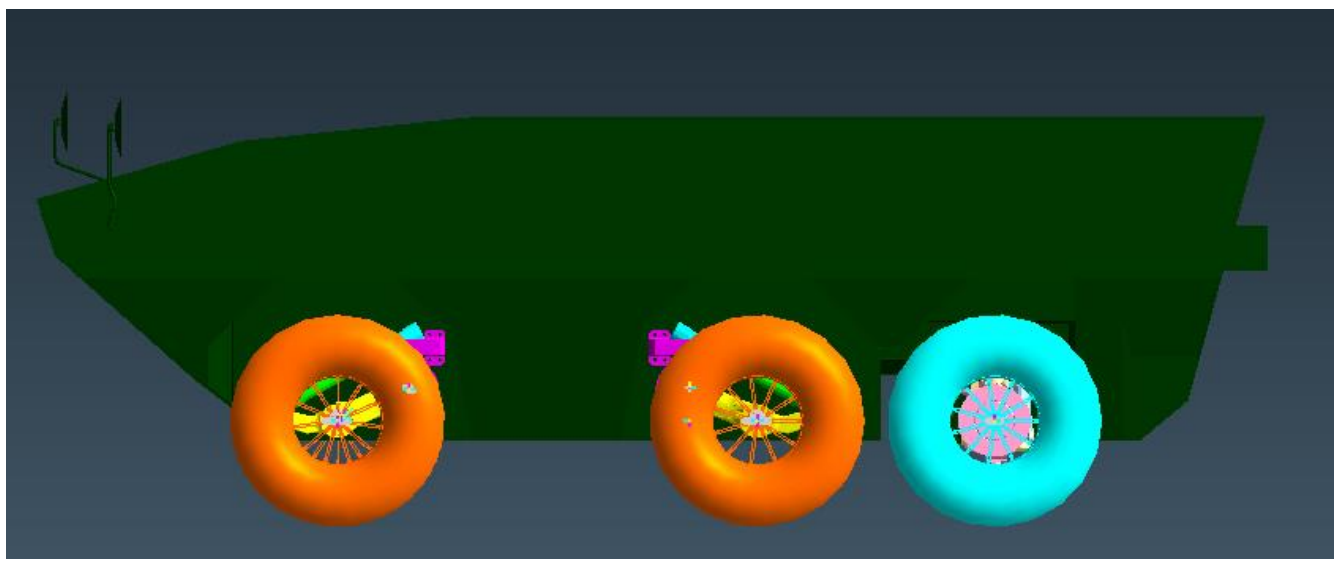

Figure 1. Vehicle model

**B. Linear model for controller design.**

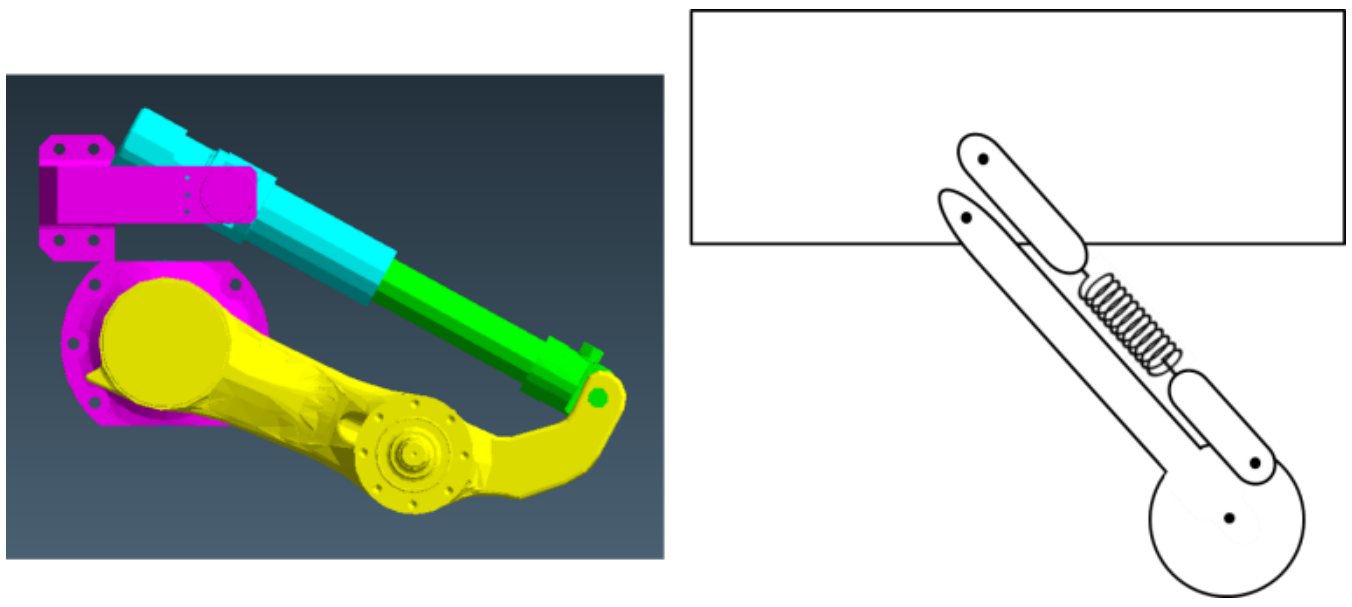

Figure 2. Suspension equivalent model

The dynamic equation is derived by solving classical action reaction force analysis. First, linearization of dynamic equations at equilibrium point was hold. However, since this approach neglects interactions between each trailing arms under the constraints of quarter-car model, this linearized model is merely accurate. To ameliorate this error, additional parameter estimation was done using online estimator and white-noise like torque input to each arms to derive nearly accurate linearized model.

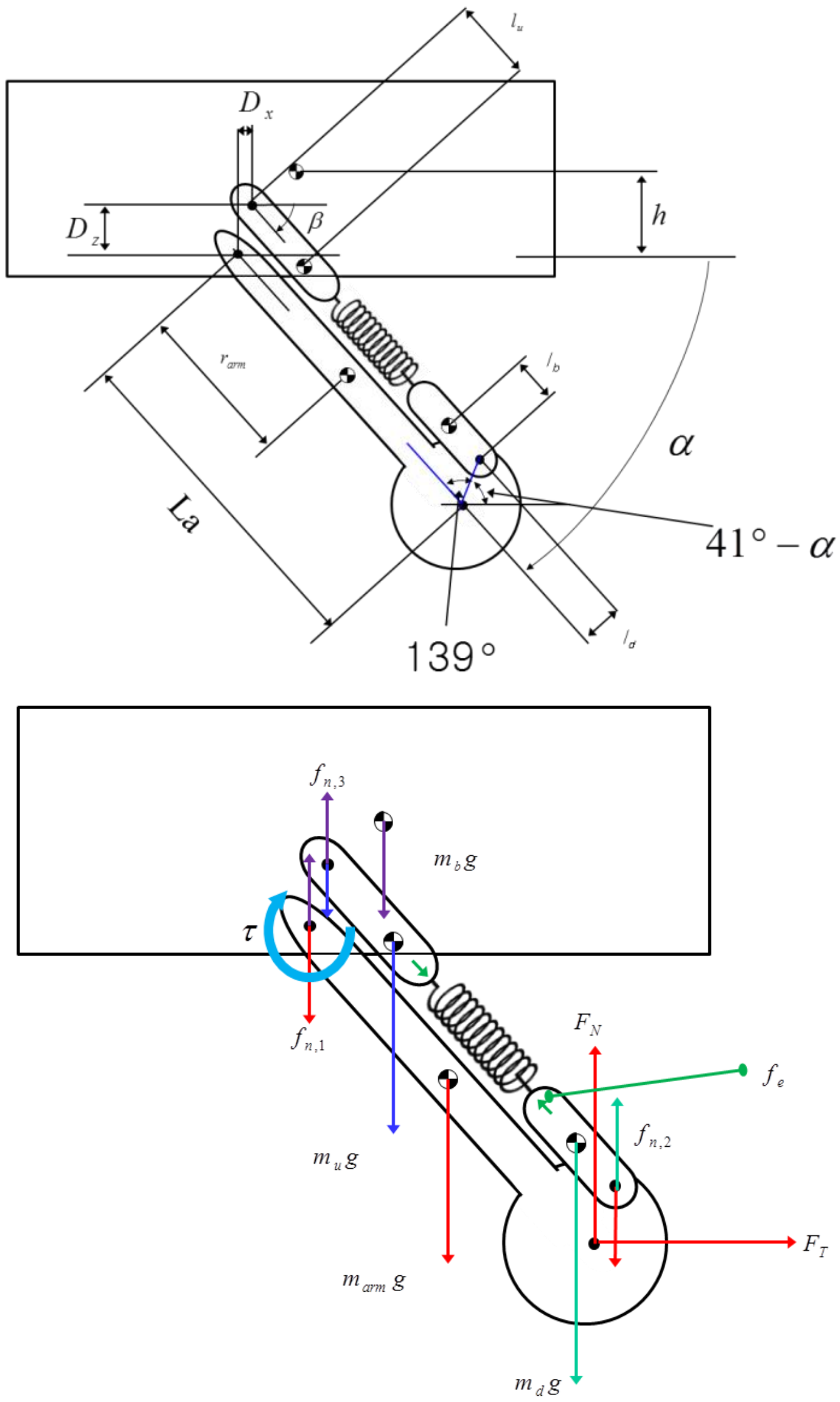

Figure 3. Equivalent model

Equation of motions of the quarter-car model is derived in (1) using action and reaction force analysis method.

$$
\begin{aligned}
& -I_{arm}\ddot{a} = -m_{arm}r_{arm}g\cdot\cos(a) + (M_{tot}\ddot{y}_b + M_{tot}g)L_a\cos(a) + \\
& L_a\cos(a)[-m_{arm}r_{arm}\{\ddot{a}\cdot\cos(a) - (\dot{a})^2\sin(a)\} - \\
& m_u l_u\{\ddot{\beta}\cdot\cos(\beta) - \sin(\beta)\cdot(\dot{\beta})^2\} - m_d L_a\{\ddot{a}\cdot\cos(a) - (\dot{a})^2\sin(a)\} \\
& + m_d l_b\{\ddot{\beta}\cdot\cos(\beta) - \sin(\beta)\cdot(\dot{\beta})^2\} - m_d l_d\{\sin(41-a)\cdot(\dot{a})^2 + \\
& \ddot{a}\cdot\cos(41-a)\}] \\
& -\{L_a\cos(a) + l_d\cos(41-a)\}\cdot[m_d < \ddot{y}_b - L_a\{\ddot{a}\cdot\cos(a) - (\dot{a})^2\sin(a)\} \\
& -l_d\{\sin(41-a)\cdot(\dot{a})^2 + \ddot{a}\cdot\cos(41-a)\} + l_b\{\ddot{\beta}\cdot\cos(\beta) - \sin(\beta)\cdot(\dot{\beta})^2 > \\
& + m_d g - f_e\sin(\beta)] - \tau
\end{aligned}
\tag{1}
$$

Define the variables $y = a$ and $u = \tau$. Then, the transfer function for each suspension system is $y(s) = \dfrac{Z_p(s)}{R_p(s)}u(s)$, where $Z_p(s), R_p(s)$ varies as vehicle's motion changes.

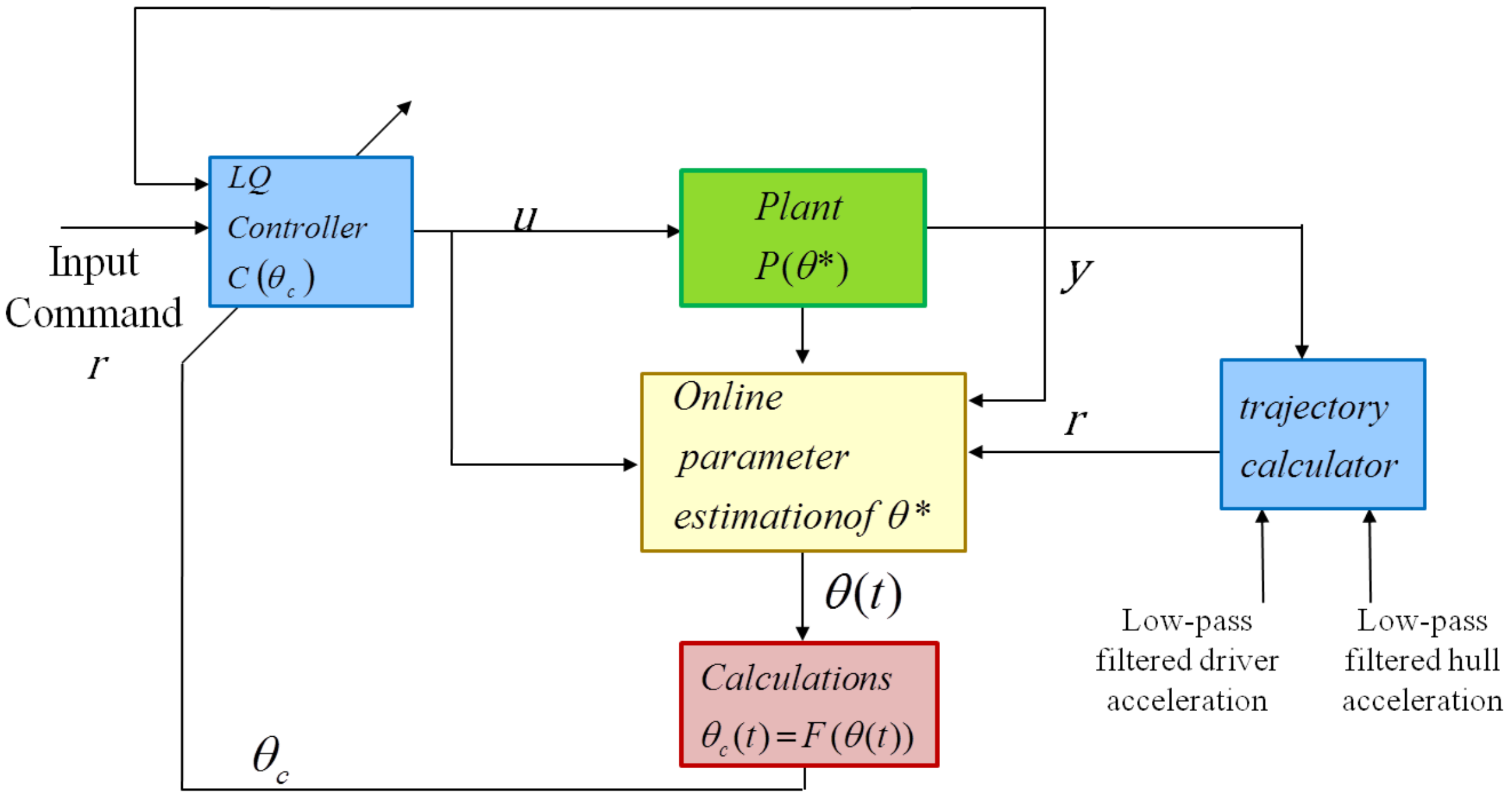

Figure 4. Adaptive LQ control diagram

## III. CONTROL LAW SYNTHESIS

### A. adaptive LQ controller design

The Adaptive LQ control schemes are shown in Figure 4. The control function *u* is designed to keep the tracking error *e*=*y*-*r* to converge to zero, where *y* is the plant output, trailing arm angle in this paper, and $r$ is reference input which will be decided from reference trajectory calculator, which discussed later. Detailed explanation of adaptive linear quadratic control scheme is given in [18]. We used gradient method as an adaptive law. The general concept of gradient method can found in [19]. A method for solving algebraic Riccati equations can be referred from [20], with detailed explanation about data abstraction for Jordan decomposition from [21, 22].

### B. Amending LQ controller for dynamic reference change

If the reference value is initially chosen (or reference equation), with proper setting in controller part [18], the plant output will smoothly track the given reference value (or equation). But if change in reference value occurs during motion, undesired protrusion leads the output to opposite direction significantly. It is because steep change in state values, resulted from abrupt change in reference value. This eccentric tendency is shown in Figure 5. This eccentricity is assuaged by, first, estimating three additional state vectors with three different references; maximum possible reference ( $r_{\max}$ ), minimum possible reference ( $r_{\min}$ ) and middle value between them ( $r_{mid}$ ). Then, a new algorithm is added that whenever change in desired reference occurs, previous state vector is disregarded and newly estimated state vector, which estimates a state vector calculated for the new reference, is replaced. This new state vector is estimated by properly weighting three different state vectors in parallel, as can be seen in (2) and (3).

$$F(x_1, x_2, x_3) = \begin{cases} \dfrac{x_3(r - r_{mid})}{r_{\max} - r_{mid}} + \dfrac{x_2(r_{\max} - r)}{r_{\max} - r_{mid}} & (r_{mid} \le r \le r_{\max}) \\ \dfrac{x_1(r_{mid} - r)}{r_{mid} - r_{\min}} + \dfrac{x_2(r - r_{\min})}{r_{mid} - r_{\min}} & (r_{\min} \le r \le r_{mid}) \end{cases} \tag{2}$$

$$\hat{x}(t + \Delta t) = \begin{cases} F(x_1(t), x_2(t), x_3(t)) & (reference\ changes) \\ \hat{x}(t) + \Delta t \cdot \{(A + K_c B)(\theta_c(t))\hat{x}(t) - K_o(\theta_c(t))(C\hat{x}(t) - e(t))\} & (reference\ fixed) \end{cases} \tag{3}$$

where $e_{\min} = y - r_{\min}, e_{\max} = y - r_{\max}, e_{mid} = y - r_{mid}$ The result of this new method is plotted in Figure 6.

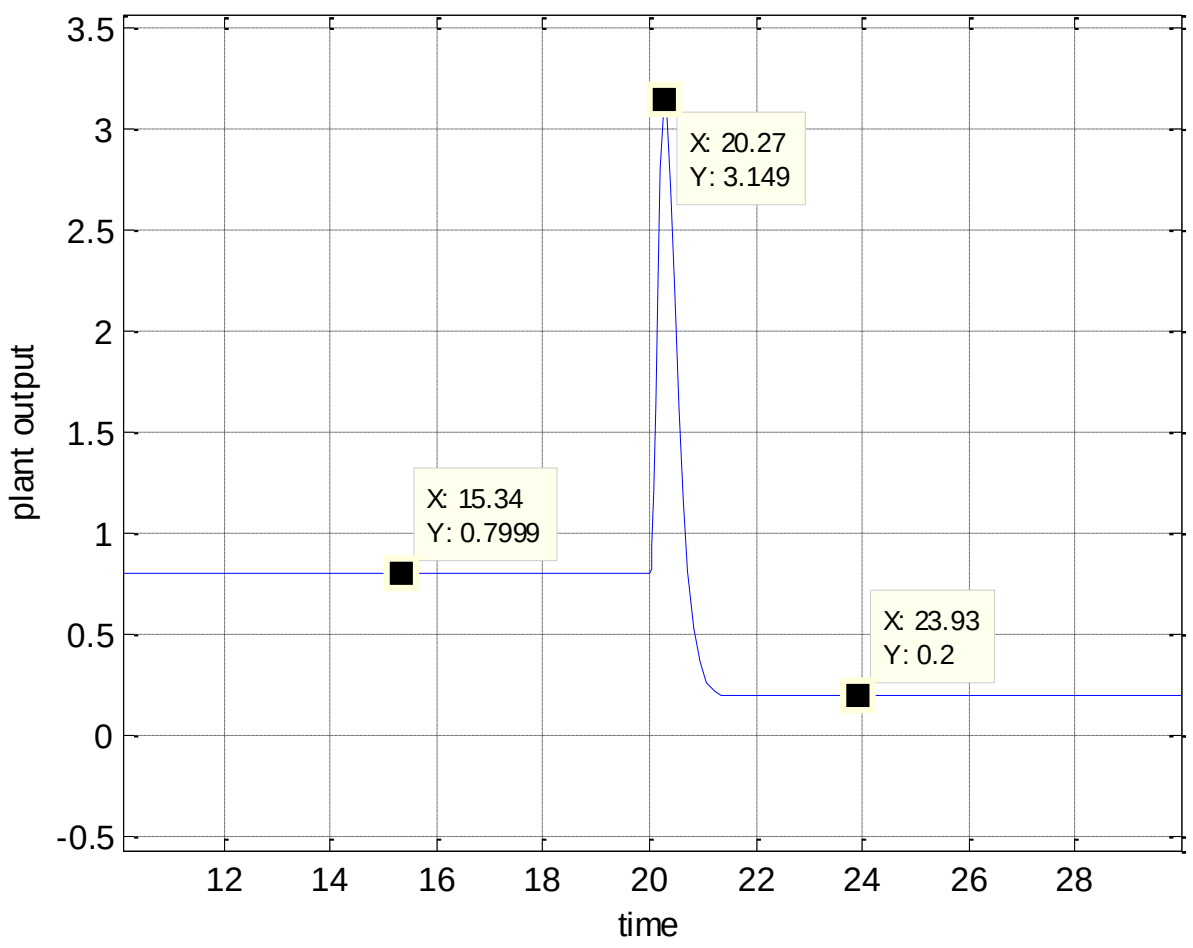

Figure 5. Reference change from 0.8 to 0.2(ordinary ALQ)

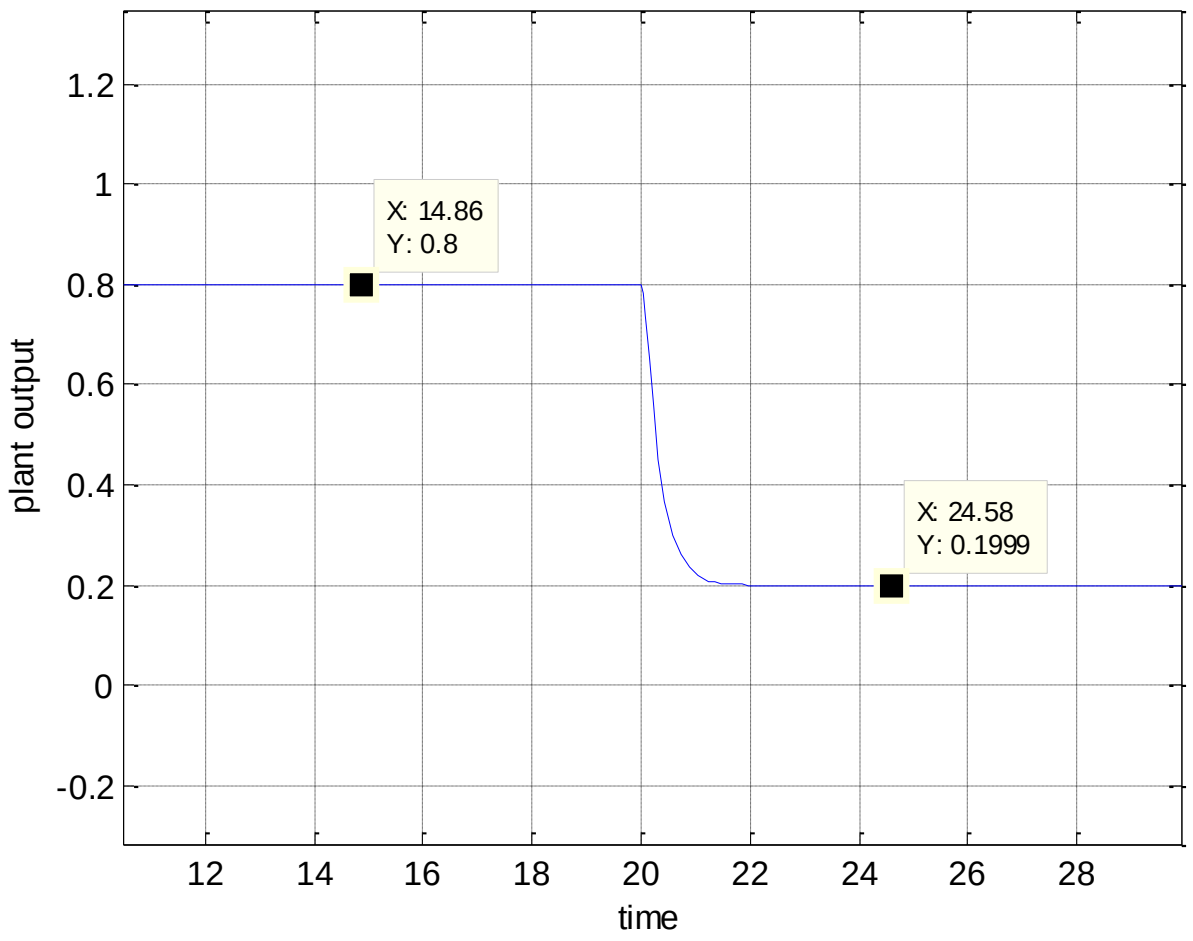

Figure 6. Reference change from 0.8 to 0.2(ameliorated)

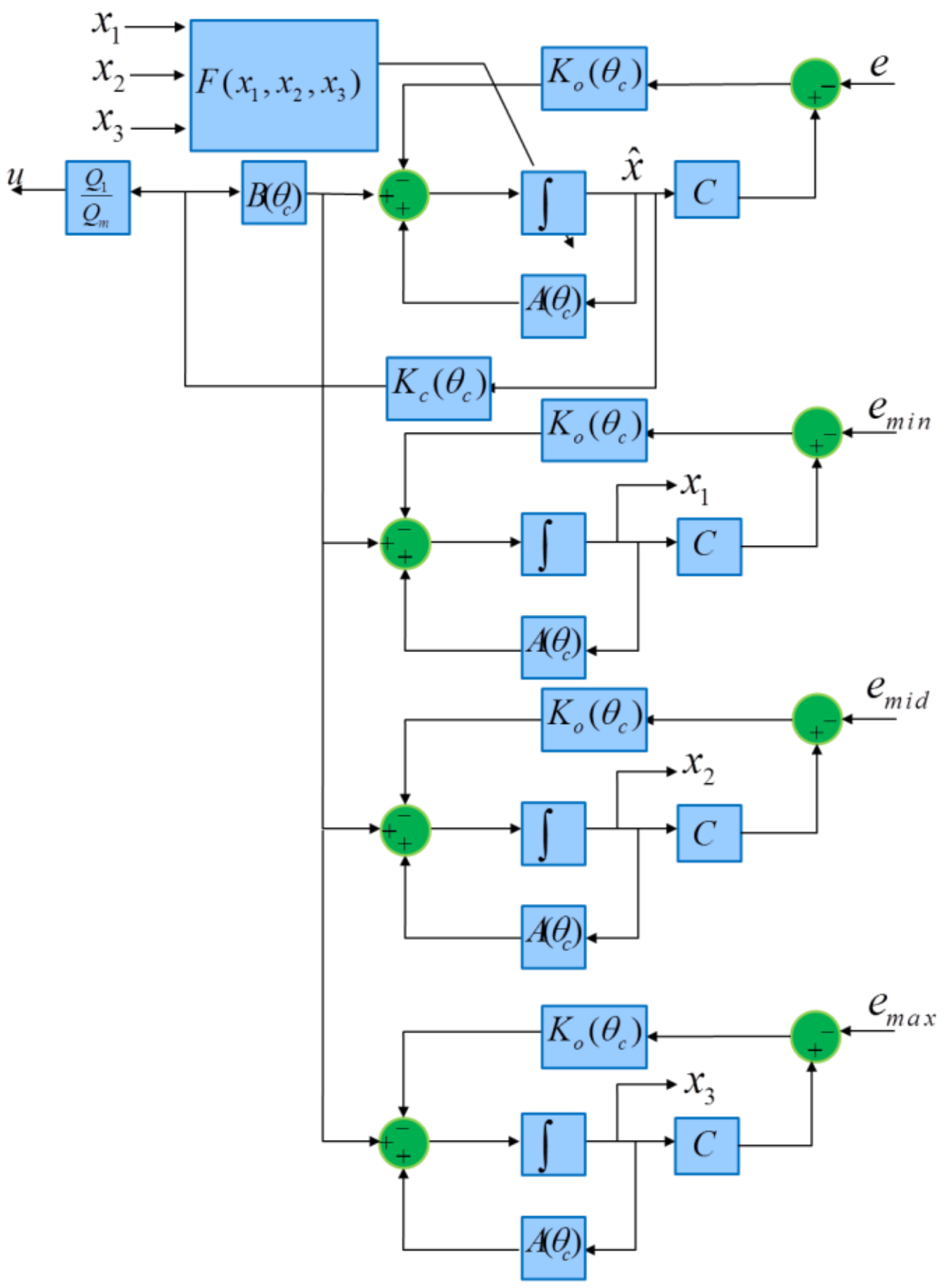

Figure 7. Ameliorated Adaptive LQ controller-part with dynamic reference allocation

It is assumed that, for any value in reference in the domain set before, the states are linearly predictable, which is similar to the concept of Taylor expansion. Scheme of suggested new method is presented in Figure 7, which corresponds to ‘LQ controller’ part of Adaptive LQ diagram in Figure 4.

## C. Reference trajectory

Previous research on suspension control, which used heave sensor, adopted adaptive back-stepping controller with reference trajectory, such as desired heave displacement and heave acceleration, etc., which is designed properly combining the vehicle’s corresponding accelerations [23]. In this paper, the controller is designed to follow somewhat similar but less complicated trajectory. Our trajectory refers to low pass filtered, in order to compensate for noise like component in road condition, vertical component of both driver acceleration and hull acceleration. Vertical accelerations due to roll motion of the vehicle is neglected in here.

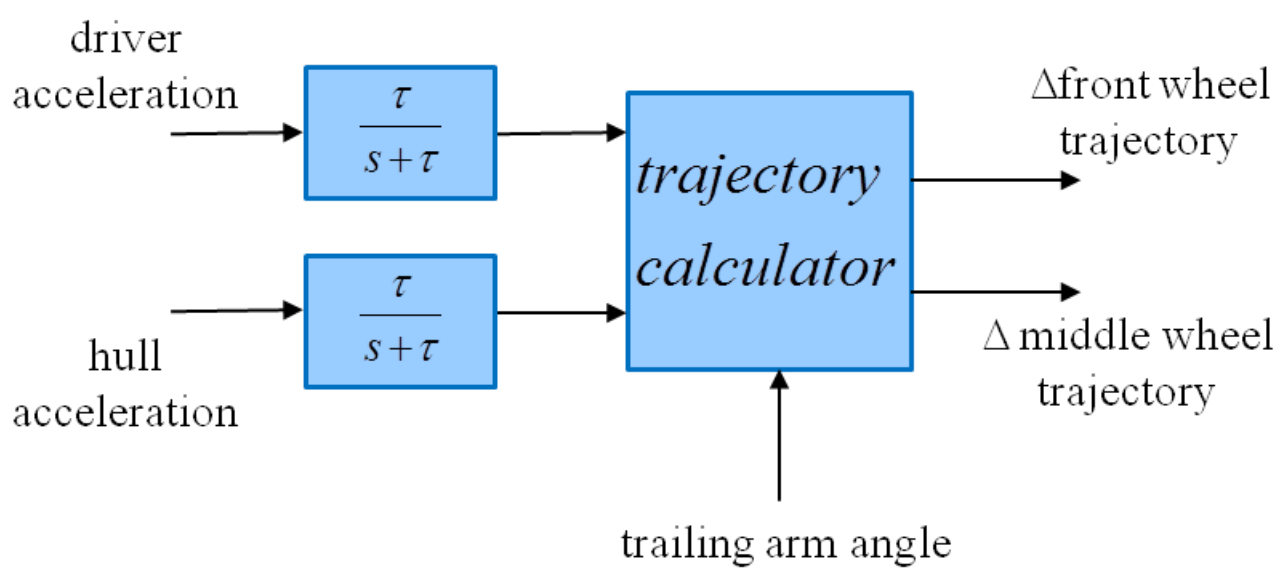

Figure 8. reference trajectory calculator

$$\tau = cutoff\ \ frequency$$
$$\ddot{y}_{driver} = driver\, acceleration$$
$$\ddot{y}_{hull} = hull\, acceleration$$
$$r_{\Delta front} = \Delta front\, wheel\ trajectory$$
$$r_{\Delta mid} = \Delta middle\, wheel\ trajectory$$
$$r_{i_fr} = initial\ reference\, value\, of\ \ front\, wheel$$
$$r_{i_m} = initial\ reference\, value\, of\ middle\, wheel$$
$$y = trailing\, arm\, angle$$
$$r_{\Delta front} = \tanh(\frac{-\ddot{y}_{driver}}{1.5}) \times \min(r_{\max} - r_{i_fr}, r_{i_fr} - r_{\min})$$
$$r_{\Delta mid} = \tanh(\frac{-\ddot{y}_{hull}}{1.5}) \times \min(r_{\max} - r_{i_m}, r_{i_m} - r_{\min})$$
$$r_{front} = \begin{Bmatrix} r_{i_fr} + k_1 \times r_{\Delta front} & (r_{\min} < y < r_{\max}) \\ r_{\max} & ((r_{\max} < y) \cap (r_{\max} < (r_{i_fr} + r_{\Delta front}))) \\ r_{\min} & ((y < r_{\min}) \cap ((r_{i_fr} + r_{\Delta front}) < r_{\min})) \end{Bmatrix}$$
$$r_{mid} = \begin{Bmatrix} r_{i_m} + k_2 \times r_{\Delta mid} & (r_{\min} < y < r_{\max}) \\ r_{\max} & ((r_{\max} < y) \cap (r_{\max} < (r_{i_m} + r_{\Delta mid}))) \\ r_{\min} & ((y < r_{\min}) \cap ((r_{i_m} + r_{\Delta mid}) < r_{\min})) \end{Bmatrix} \quad (4)$$

The cut off value $1.5m/s^2$ is set in both $r_{\Delta front}$ and $r_{\Delta mid}$, since that much acceleration in either hull or driver position is already very undesirable. $k_1, k_2$ are simple constants. Higher $k_1, k_2$ provides greater reaction to hull and driver acceleration.

### D. Stability analysis

Stability of a general adaptive LQ controller is proven in [18]. In order to prove the stability of the system, we should prove the stability of $\hat{x}$ no matter how frequently reference change required. Since the stability of $\hat{x}$ without reference change is guaranteed, if $F(x_1, x_2, x_3)$ is always stable, which means $x_1$, $x_2$, and $x_3$ is always stable, then the system is stable. First, examining the case right after the reference change is required:

$$\begin{aligned} \dot{x}_k &= Ax_k + BK_c\hat{x} - K_o(Cx_k - \{y - (r + \Delta_k)\}) \quad (k = 1,2,3) \\ \dot{\hat{x}} &= A\hat{x} + BK_c\hat{x} - K_o(C\hat{x} - (y - r)) \end{aligned} \tag{5}$$

Subtracting these two equations in (5):

$$(\dot{\hat{x}} - \dot{x}_k) = (A - K_oC)(\hat{x} - x_k) + K_o\Delta_k \tag{6}$$

Since $A - K_oC$ will set to be stable and $\hat{x}$ converges to zero, $x_k$ is stable, and converges to $K_o\Delta_k$.

Furthermore, after a change in desired reference, the error between linearly predicted states is same as the state for the new reference:

$$\begin{aligned} \dot{x}_{ref} &= Ax_{ref} + BK_c\hat{x} - K_o(Cx_{ref} - \{y - (r + \Delta_{ref})\}) \\ (\dot{\hat{x}} - \dot{x}_{ref}) &= (A - K_oC)(\hat{x} - x_{ref}) + K_o\Delta_{ref} \\ (\dot{\hat{x}} - \dot{x}_1) &= (A - K_oC)(\hat{x} - x_1) + K_o\Delta_1 \\ (\dot{\hat{x}} - \dot{x}_2) &= (A - K_oC)(\hat{x} - x_2) + K_o\Delta_2 \end{aligned} \tag{7}$$

If $\Delta_{ref} = c_1\Delta_1 + c_2\Delta_2 \qquad (c_1 + c_2 = 1)$

$$\begin{aligned} (c_1\dot{\hat{x}} - c_1\dot{x}_1) &= (A - K_oC)(c_1\hat{x} - c_1x_1) + K_oc_1\Delta_1 \\ (c_2\dot{\hat{x}} - c_2\dot{x}_2) &= (A - K_oC)(c_2\hat{x} - c_2x_2) + K_oc_2\Delta_2 \\ \{\dot{\hat{x}} - (c_1\dot{x}_1 + c_2\dot{x}_2)\} &= (A - K_oC)(\hat{x} - (c_1x_1 + c_2x_2)) + K_o(c_1\Delta_1 + c_2\Delta_2) \end{aligned} \tag{8}$$

Let $c_1x_1 + c_2x_2 = \hat{x}_{ref} = F(x_1, x_2, x_3)$,

$$(\dot{\hat{x}} - \dot{\hat{x}}_{ref}) = (A - K_oC)(\hat{x} - \hat{x}_{ref}) + K_o\Delta_{ref} \tag{9}$$

where $x_{ref}$ is a state for the new reference that we assume being calculated prior to the reference change occurs. Same result holds for $x_2$ and $x_3$.

Since $\hat{x}_{ref} \rightarrow \Delta_{ref}$ and $x_{ref} \rightarrow \Delta_{ref}$ with same speed and same initial condition, which can be achieved by simply assuming all states starts from zero initial value, before reference change occurs, $\hat{x}$ is always guaranteed to be stable. Figure 9 gives picture of an example when $x_1 = 0.3, x_2 = 0, x_3 = -0.3$ and reference change occurs at time=1.5. $x_{hat}$ in the Figure 9 refers to $\hat{x}_{ref}$ in equation (9).

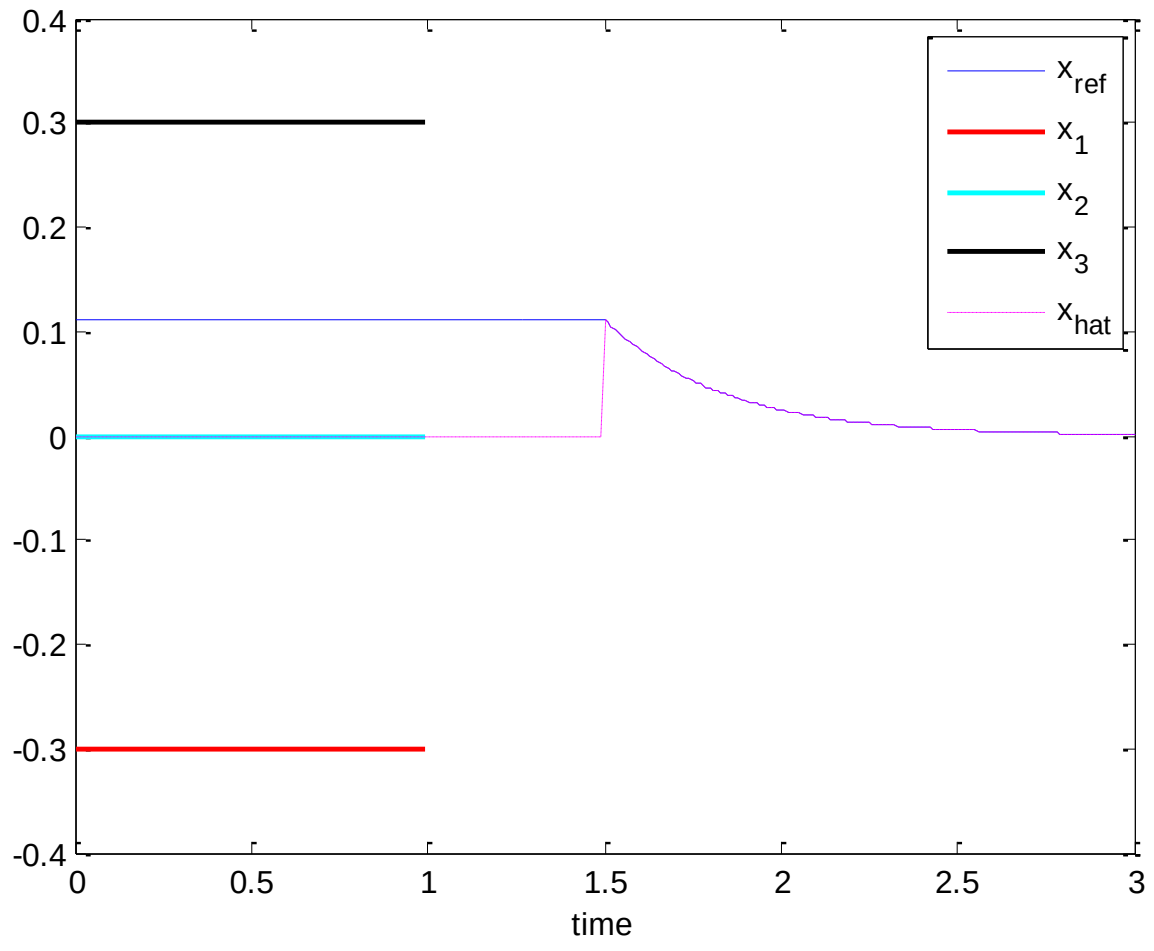

Figure 9. Reference change scheme (change at time=1.5)

## IV. Design Example

### A. Ride quality

There would be many factors that determine ride comfort of the road vehicle including road roughness, external noise, etc. Various methods of rating the severity of this kind of exposure have been suggested previously. In order to evaluate the improvement in the ride quality of this vehicle, vertical accelerations at driver's position and crew member's position will be measured and compared. This vibration will be evaluated using the criterion suggested by the International Standard Organization (ISO2631) which provides the frequency weighting curve relating to vertical acceleration [24]. The weighted RMS vertical acceleration can be computed from the equation,

$$a_{rms} = \left[ \frac{1}{T} \int_0^T a_w^2(t)\, dt \right]^{\frac{1}{2}} \qquad (10)$$

where $a_w(t)$ is the frequency weighted vertical acceleration, with frequency weightings and multiplying factors are defined in Table 3 of [25]. Table 1 shows the scale of discomfort suggested in

ISO2631.

Table 1. Acceleration level and degree of comfort (ISO2631-1, 1997).

| Acceleration level | Degree of comfort |
|---|---|
| (1) Less than $0.315m/s^2$ | Not uncomfortable |
| (2) $0.315-0.63m/s^2$ | A little uncomfortable |
| (3) $0.5-1m/s^2$ | Fairly uncomfortable |
| (4) $0.8-1.6m/s^2$ | Uncomfortable |
| (5) $1.25-2.5m/s^2$ | Very uncomfortable |
| (6)Greater than $2m/s^2$ | Extremely uncomfortable |

**B. Numerical analysis**

The amended ALQ controllers are applied to the target system with ideal actuators, without limit on torque, on the all wheels. The velocity of the vehicle at place where vertical accelerations of driver and crew member are examined is 25km/h. The simulation data are sampled at frequency of 1 KHz. The vehicle was examined on three different road conditions; bump road input, road input with limited ramp and sinusoidal road input [26]. First, time responses of the vehicle for the bump road input are shown. Bump road input is one of the most encountered road surface irregularities during actual driving situation. Bump road input in time scale can be formulated as:

$$y_0(t)=\begin{Bmatrix} \dfrac{h[1-\cos(8\pi t)]}{2}, & 0\le t\le 0.25 \\ 0 & otherwise \end{Bmatrix} \tag{11}$$

where $h$ is the peak of the bump road input (0.04 meter) for the both right and left wheels. Bump road input is depicted in Figure 10.

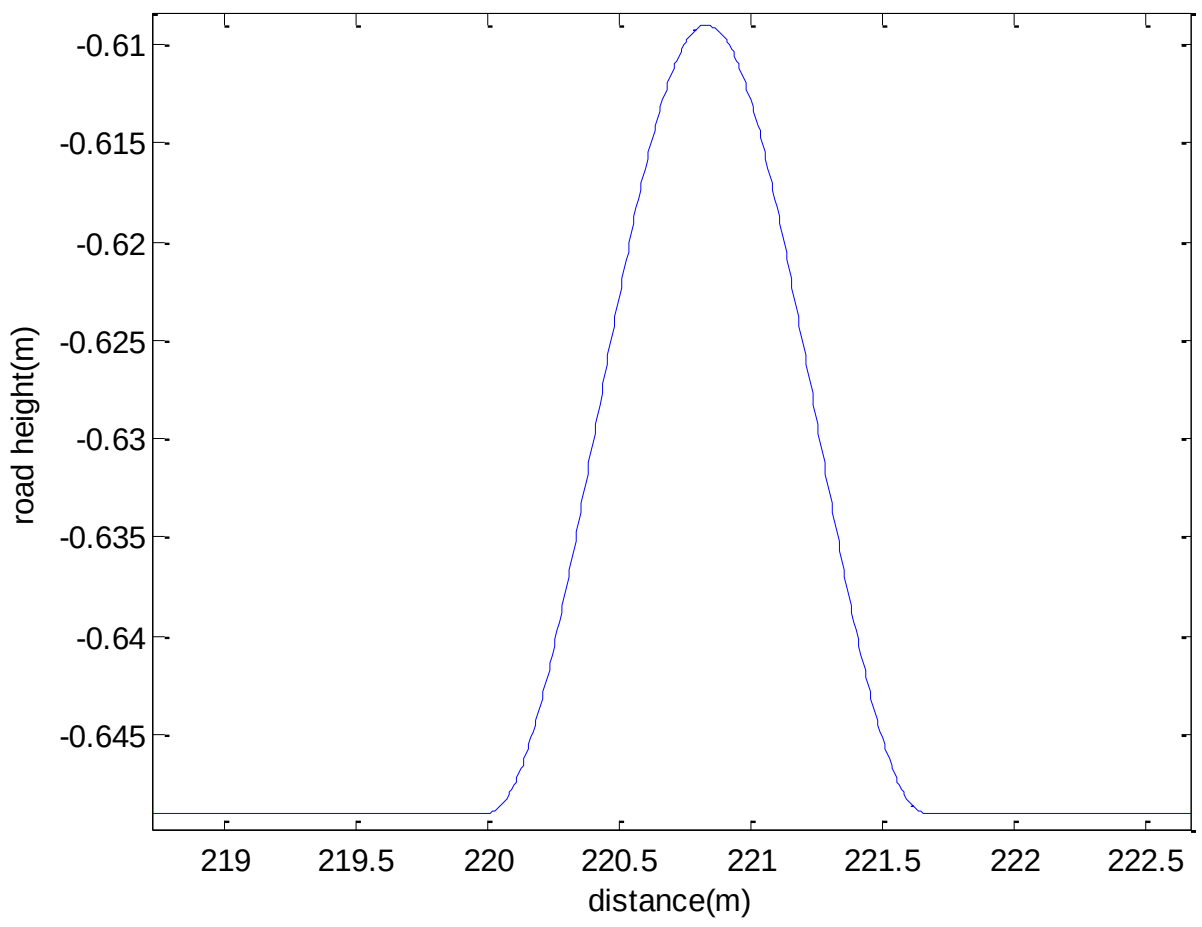

Figure 10. Bump road inputs.

Time $t$ is set to 0 when the vehicle starts to drive on the bump road. Simulation results of the heave, pitch motions, the driver's vertical position and corresponding accelerations for the bump road input are given in Figure 11. It is seen from this figure that the magnitudes for the acceleration in the driver's position is significantly decreased. Also, the displacements and accelerations for our adaptive controlled active suspensions vanish faster than the passive suspensions. These results evince the efficiency of the reference changing adaptive controller. Specifically, decreased vertical accelerations, as in Figure 11, indicate that the ride comforts are improved at driver and crew member's position. The frequency weighted RMS values for the heave, pitch and driver position's accelerations of the vehicle body are presented in Figure 12. The decrease in the RMS values is between 46.71 % and 59.05 % which shows increased ride comfort of the vehicle. Peak values for the heave, pitch and driver position's accelerations are also depicted in Figure 12. Moreover, the peak values of accelerations are decreased with suggested control method, that is, the maximum accelerations sensed by the crew members and driver are decreased between 40.35 % and 49.81 %. There is difference in initial vertical displacement of the driver between passive case and active case because $r_{i_fr}$ set the initial reference value of the front wheels' arm angle. For the passive case, it seems like road elevation occurred after bump road input. But this is because this road input result slight change in arm angles of the vehicle and, consequently, shifted equilibrium point of the vehicle due to changed geometry, as seen in Figure 13. The reference trajectory from reference calculator and corresponding variation in arm angle is presented in Figure 14. It is clear that these reference values give the controller the tendency of the control, although it is actually demanding controller to exactly follow its given reference, which is almost impossible due to insufficient time to follow.

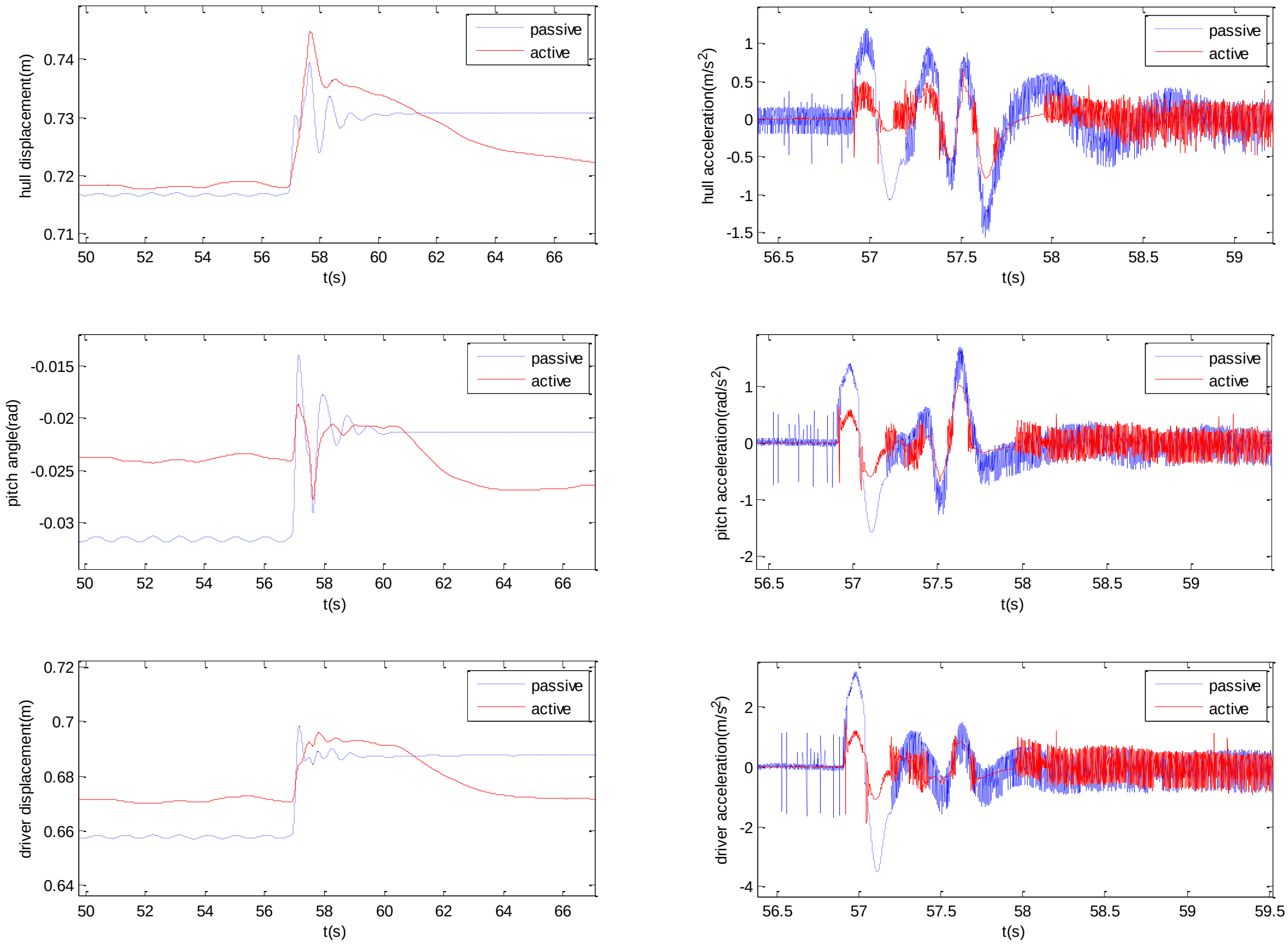

Figure 11. Heave, pitch and driver position's vertical motion of the sprung mass and related accelerations for bump road input

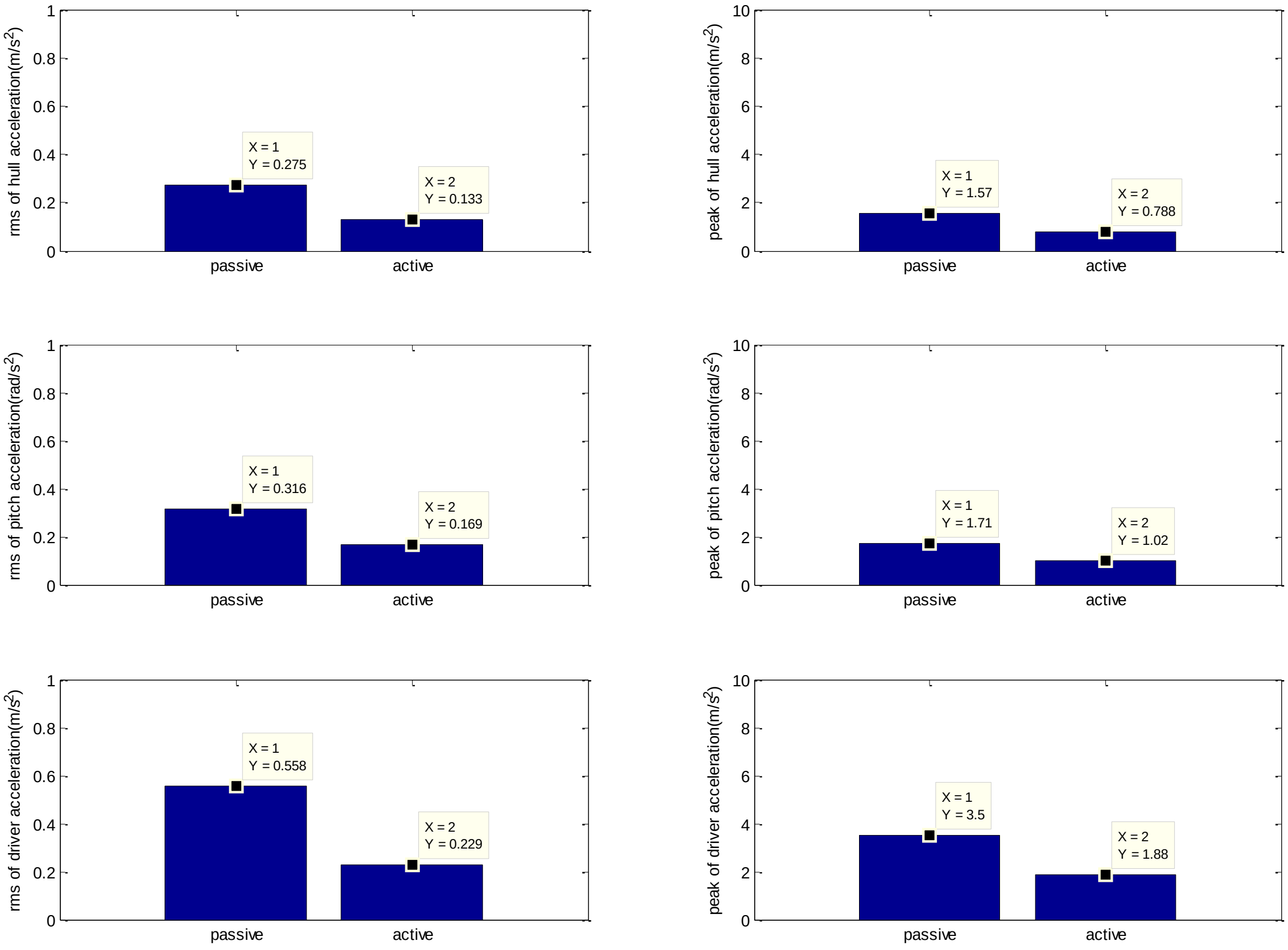

Figure 12. RMS and peak values for the heave, pitch and driver accelerations for bump road input

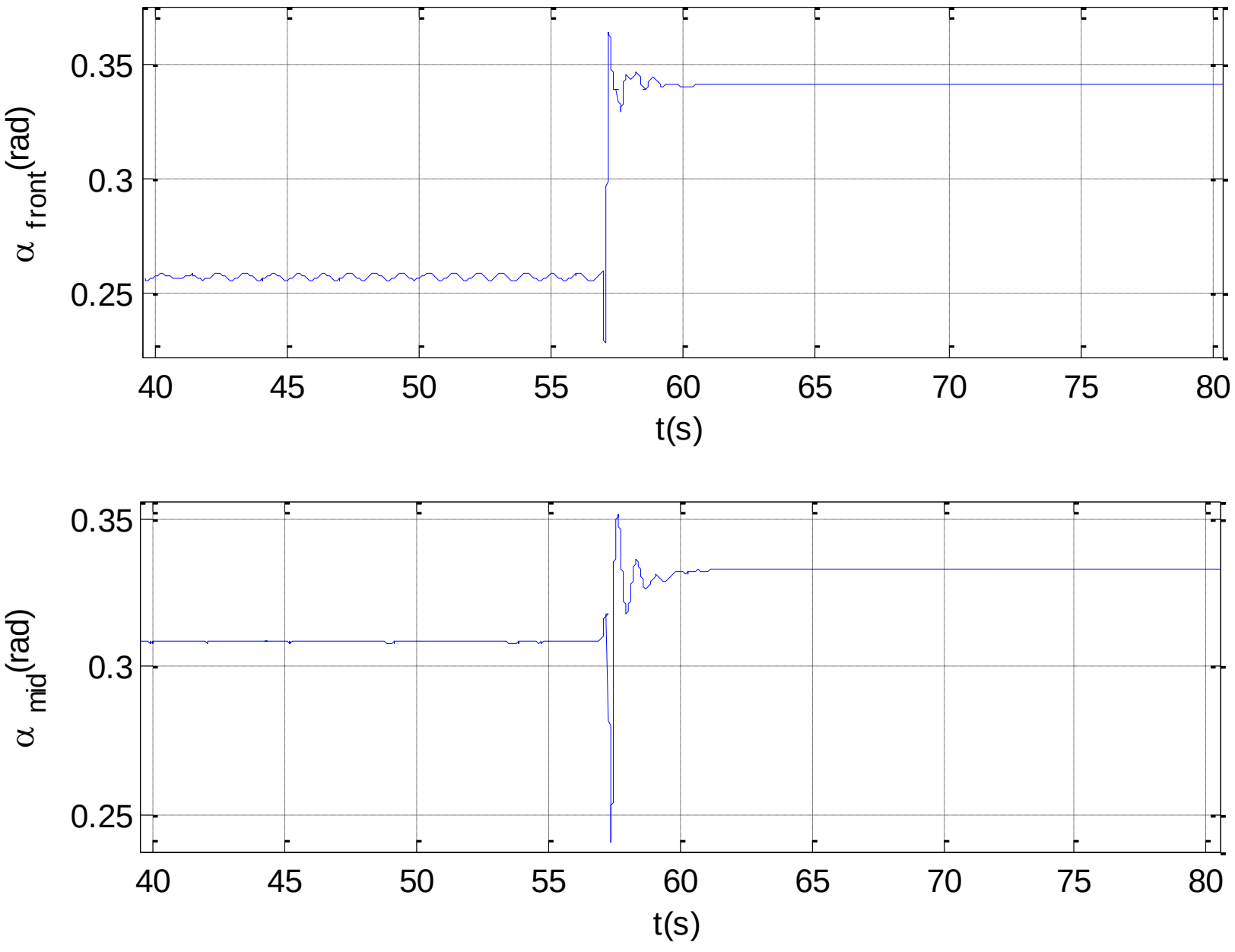

Figure 13. Suspension arm angle of the vehicle on bump road input

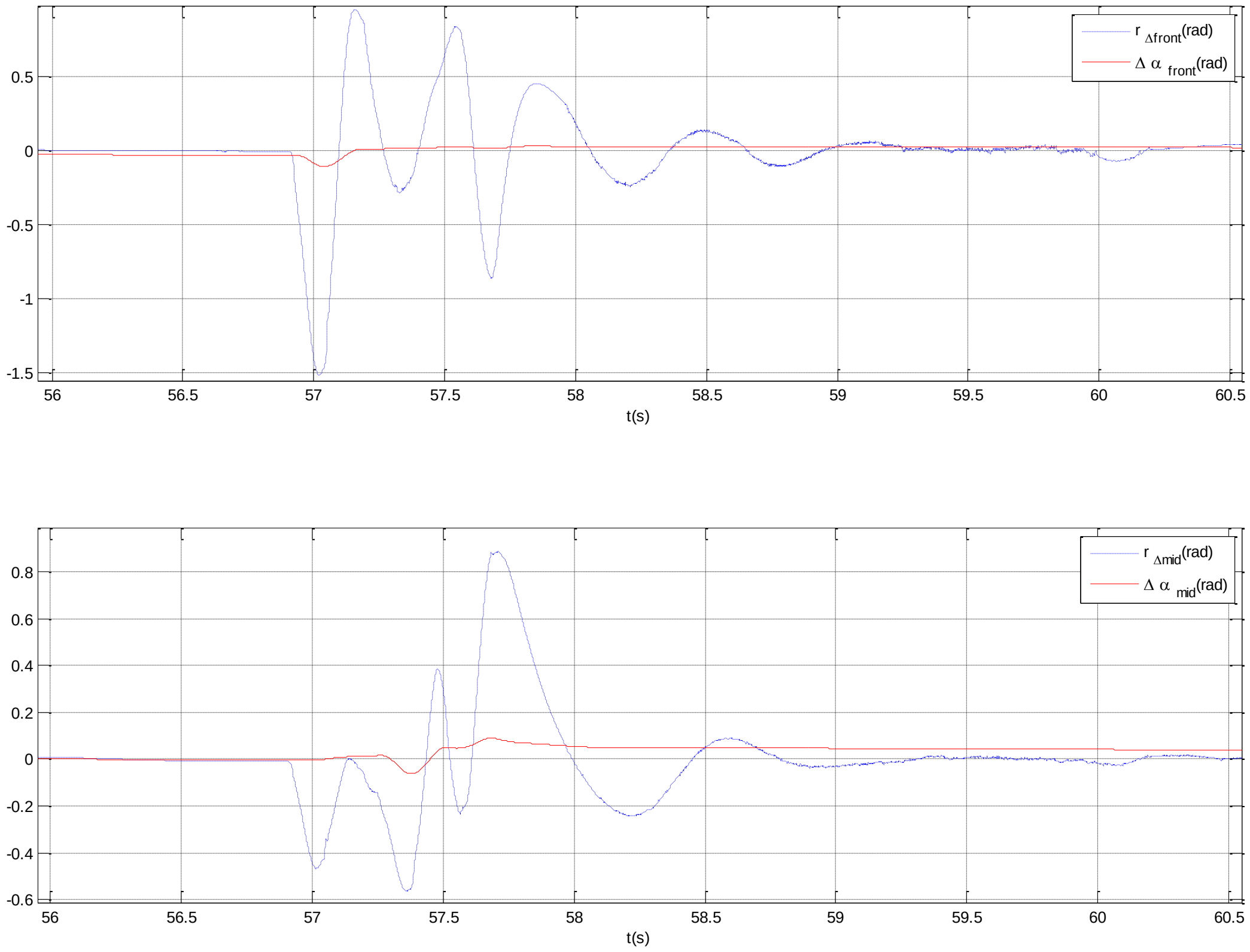

Figure 14. Reference trajectory and corresponding actual arm angle on bump road input

Next, the vehicle was simulated on a road input with limited ramp, Figure 15. This type of road inputs usually used to test the performance of the suspension when the vehicle faces a sudden change in the road surface elevation [27]. This type of road input also tests whether the suspension under suggested controller preserves its working space since the inherent suspension working space of the vehicle would not be changed while pushing up or pulling down the vehicle body in order to compensate for the change in road elevation. The road input with limited ramp is formulated as below where h is the final road surface elevation.

$$y_{0(t)} = \begin{cases} 0, & 0 \le t < 1 \\ 10h(t-1), & 1 \le t < 1.1 \\ h, & 1.1 \le t \end{cases} \qquad (12)$$

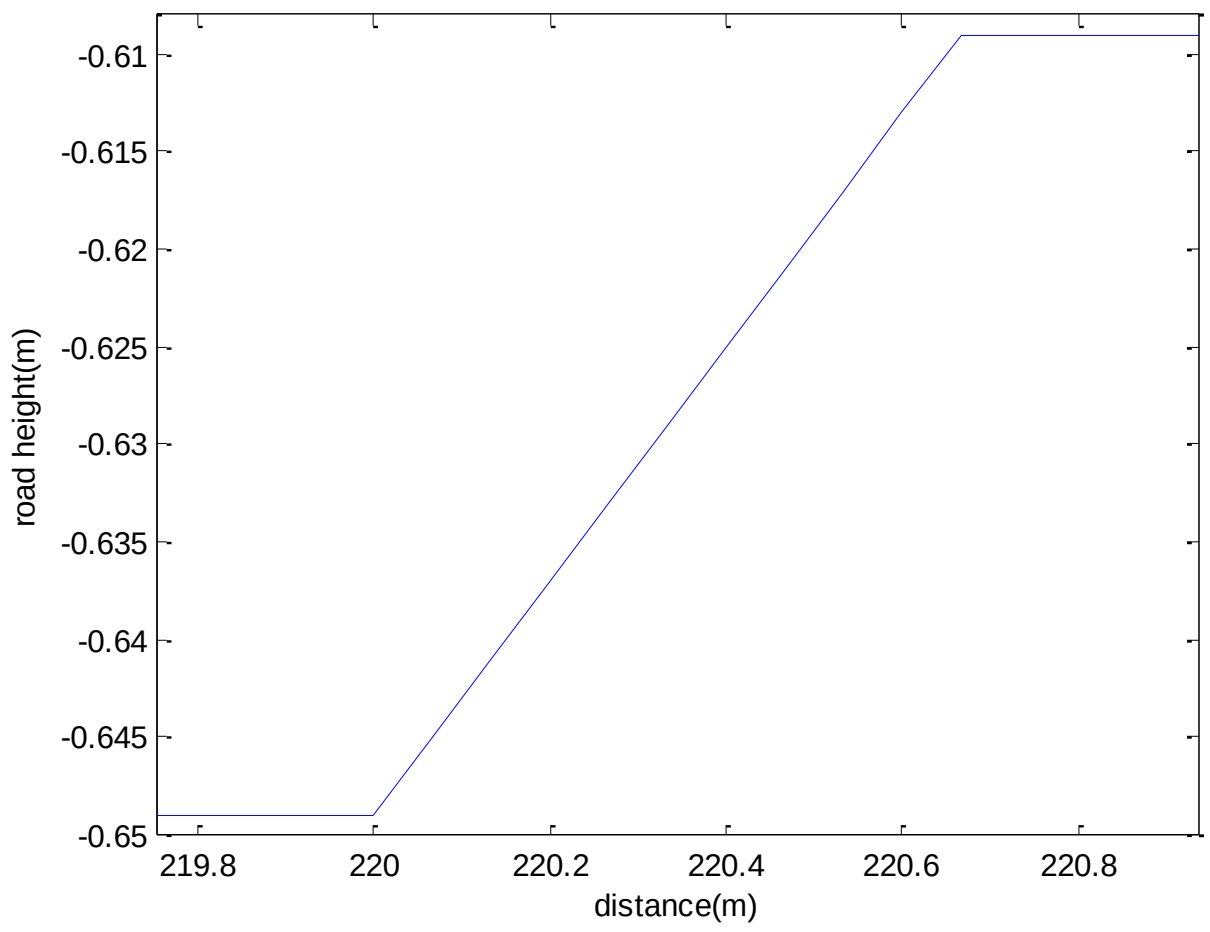

Figure 15. Road input with limited ramp

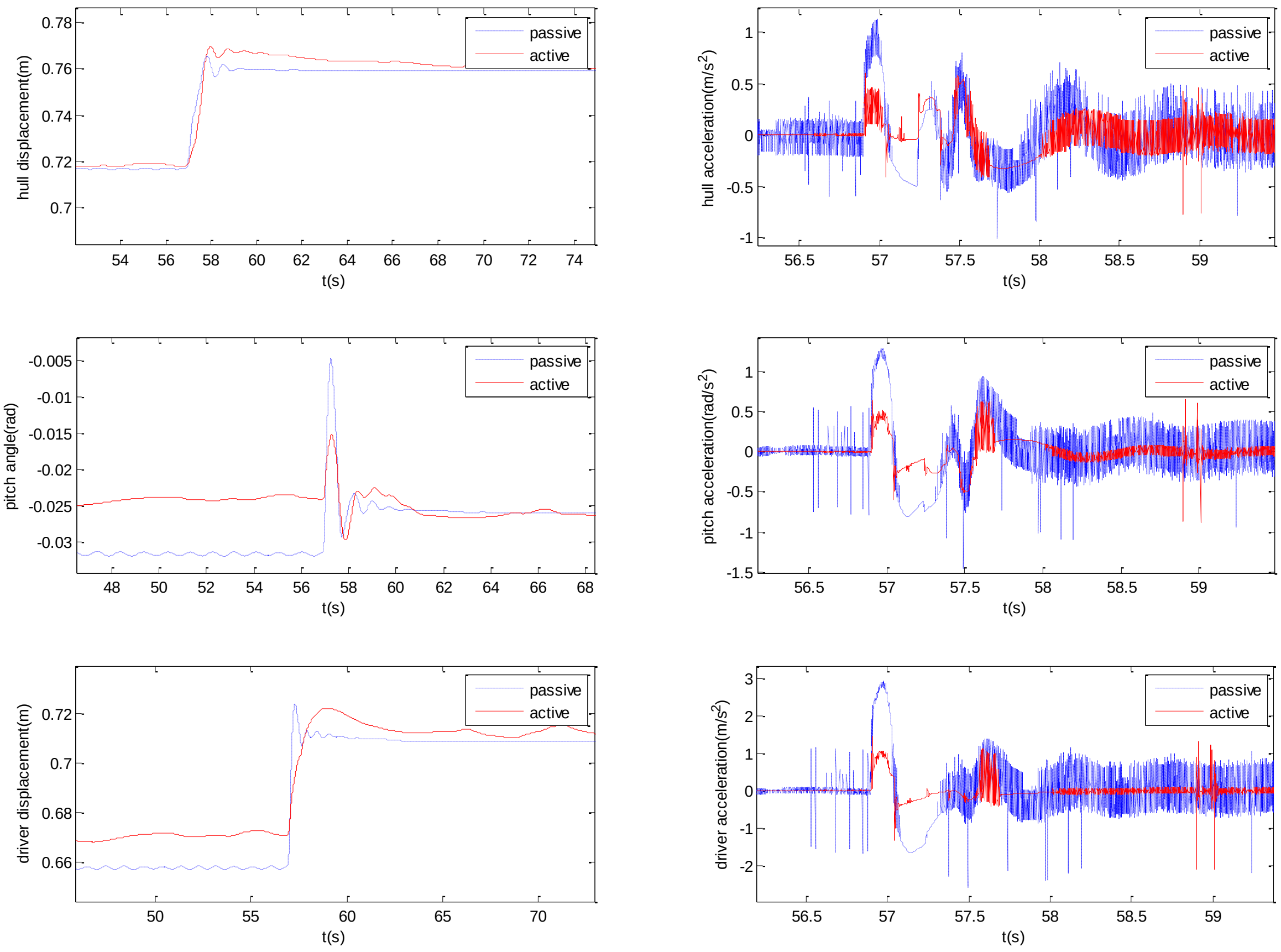

Figure 16. Time histories for the vehicle body heave, pitch and driver's vertical motion and corresponding accelerations for the road input with limited ramp

Simulation result for the vehicle body heave, pitch, driver's vertical motions and corresponding accelerations for the road input with limited ramp are presented in Figure 16. This figure clearly shows that the target vehicle reaches the final height value of the road input with improved ride comfort. Suspension deflections in both front and mid arm of the vehicle for this road input are given in Figure 17. It is observed from this figure the suspension space needed to operate is preserved.

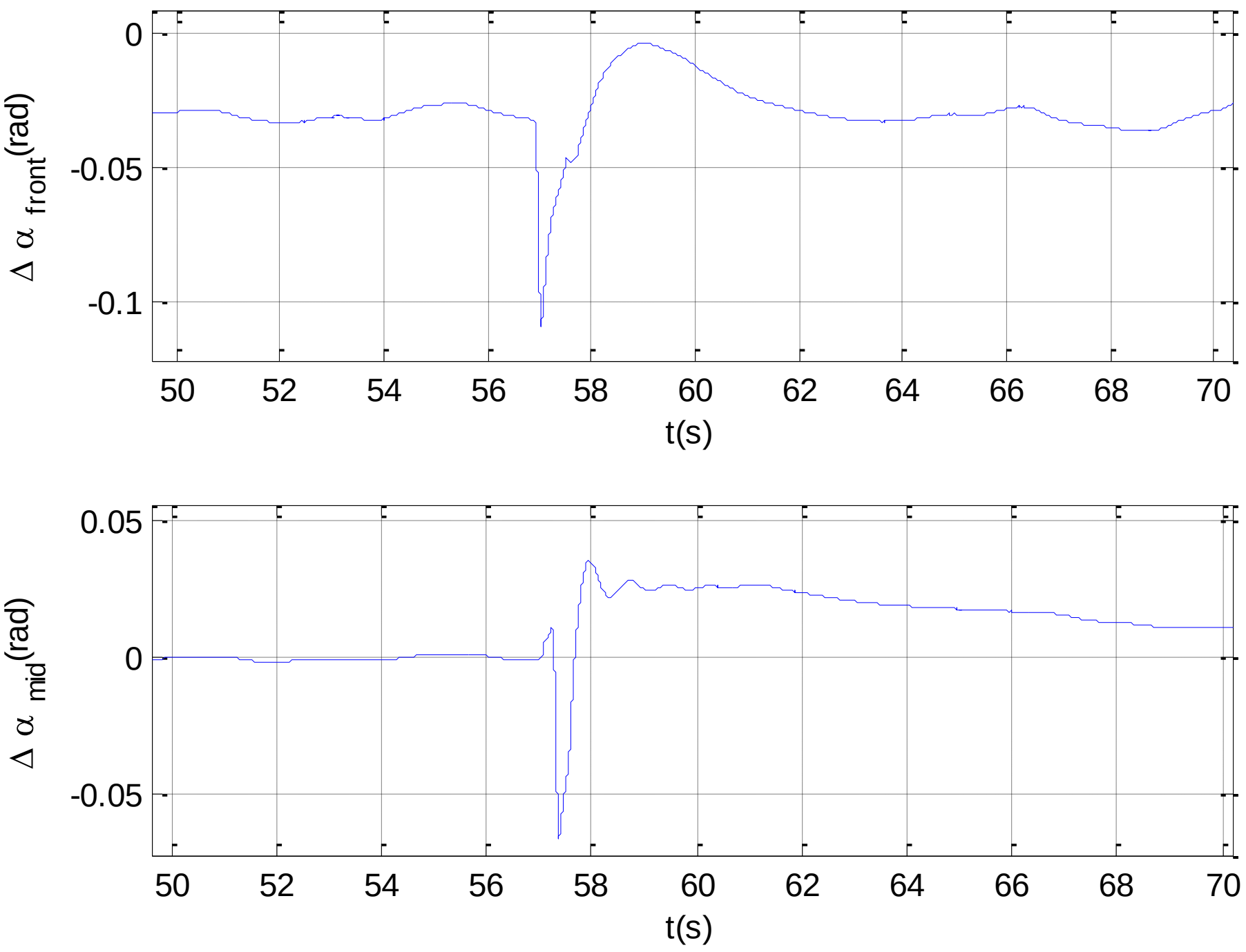

Figure 17. Variation of the suspension deflections for the road input with limited ramp

Figure 18 presents the weighted RMS values for the heave, pitch and driver's position accelerations of the sprung mass for the road input with limited ramp. The ride comfort is improved since the RMS values for the heave, pitch and driver accelerations of the sprung mass are decreased between 39.41 % and 65.09 %. The peak values for the heave, pitch and driver accelerations of the sprung mass are also shown in Figure 18. The reference trajectory from reference calculator and corresponding variation in arm angle is presented in Figure 19.

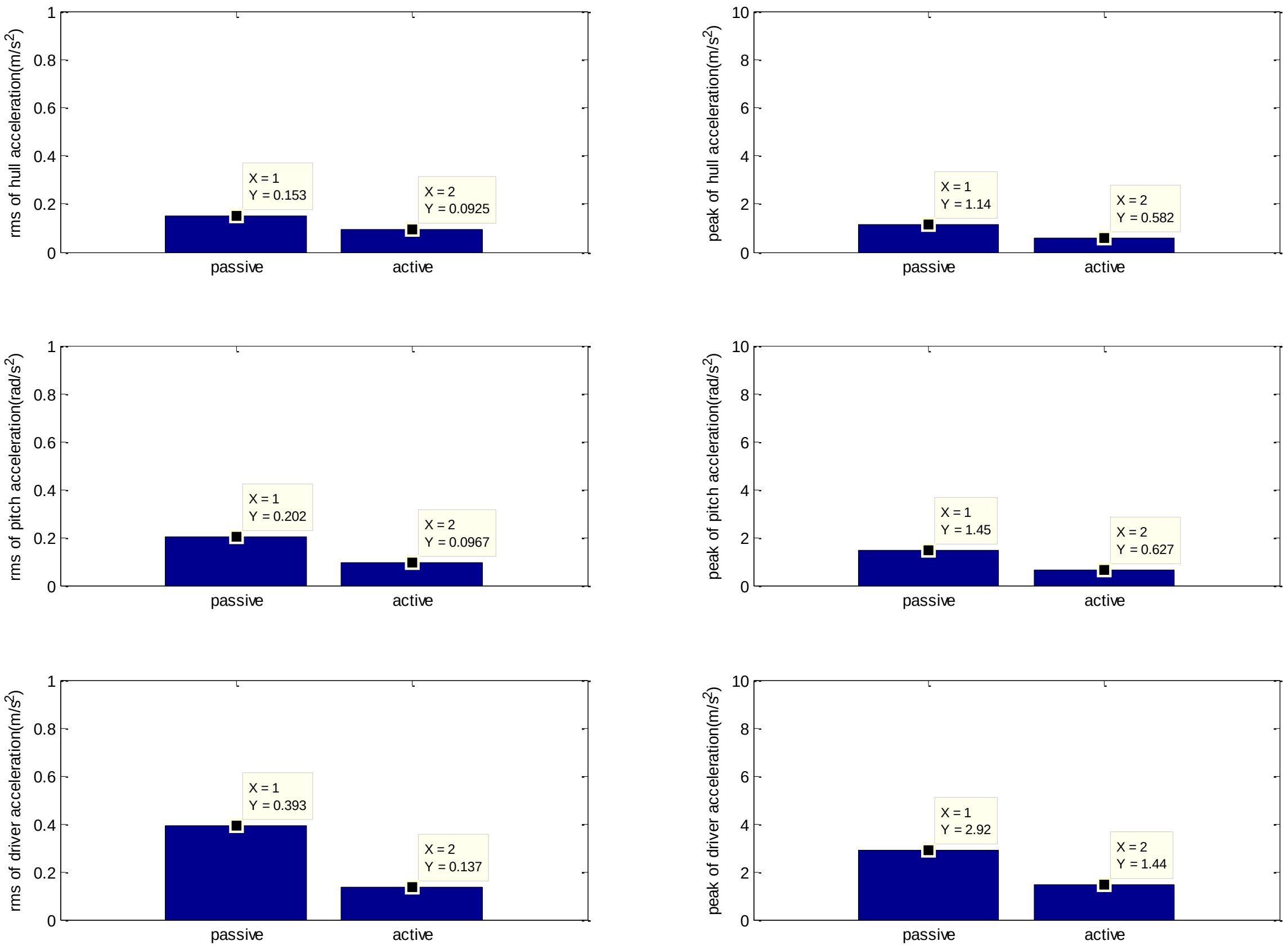

Figure 18. The RMS and peak values for the heave, pitch and driver accelerations for the road input with limited ramp

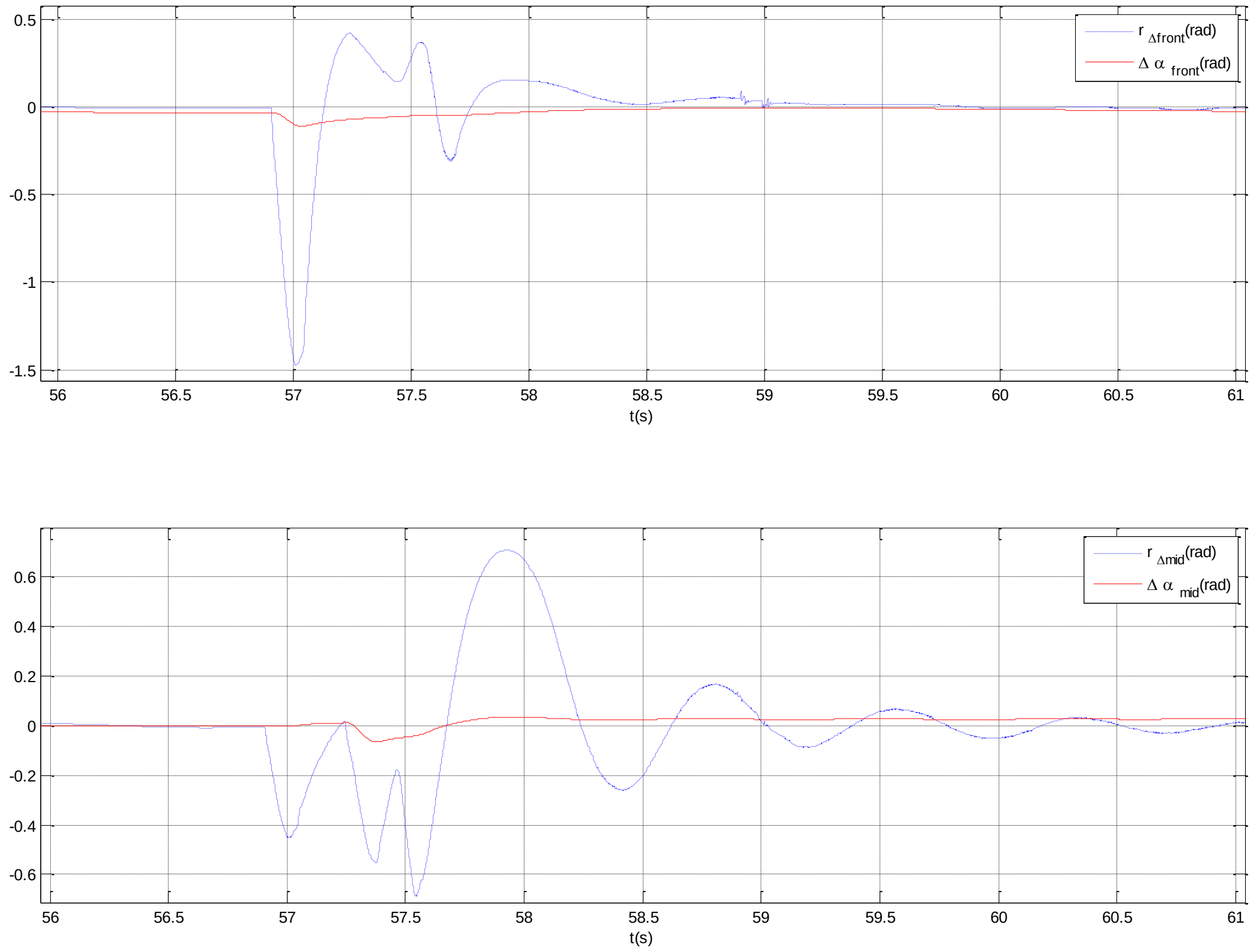

Figure 19. Reference trajectory and corresponding actual arm angle on road type with limited ramp

Finally, for the time domain analysis, a sinusoidal road input is applied to the full vehicle model in order to test the performance of the controller for severe periodic road conditions and to see the capability of the controller keep a stable and comfortable reference value. The sinusoidal road input to the front wheels is formulated as below with h being the magnitude of the road input:

$$y_0(t) = \frac{h}{2}(1+\sin(\pi t - \frac{\pi}{2})) \qquad (h = 0.04) \tag{13}$$

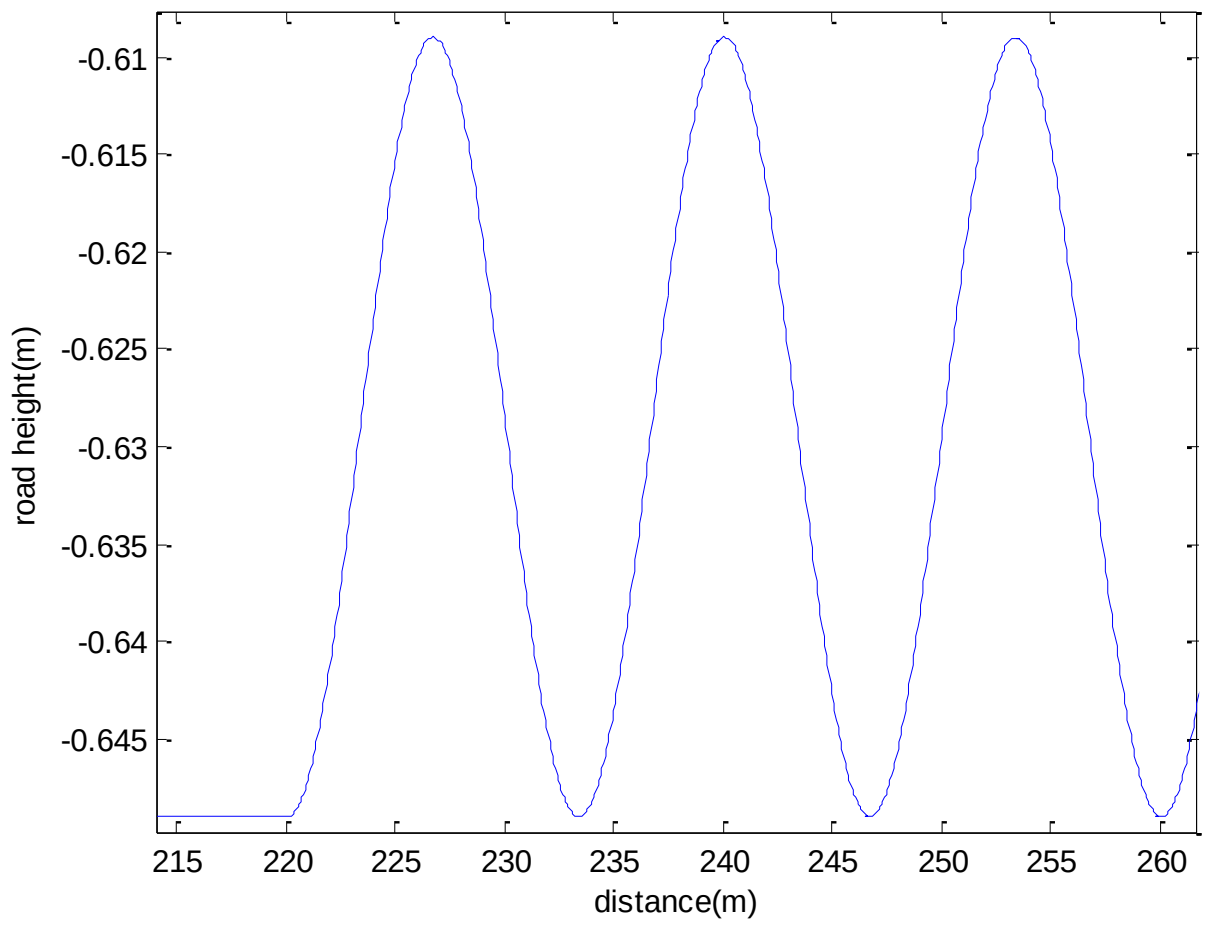

Figure 20. Sinusoidal road input

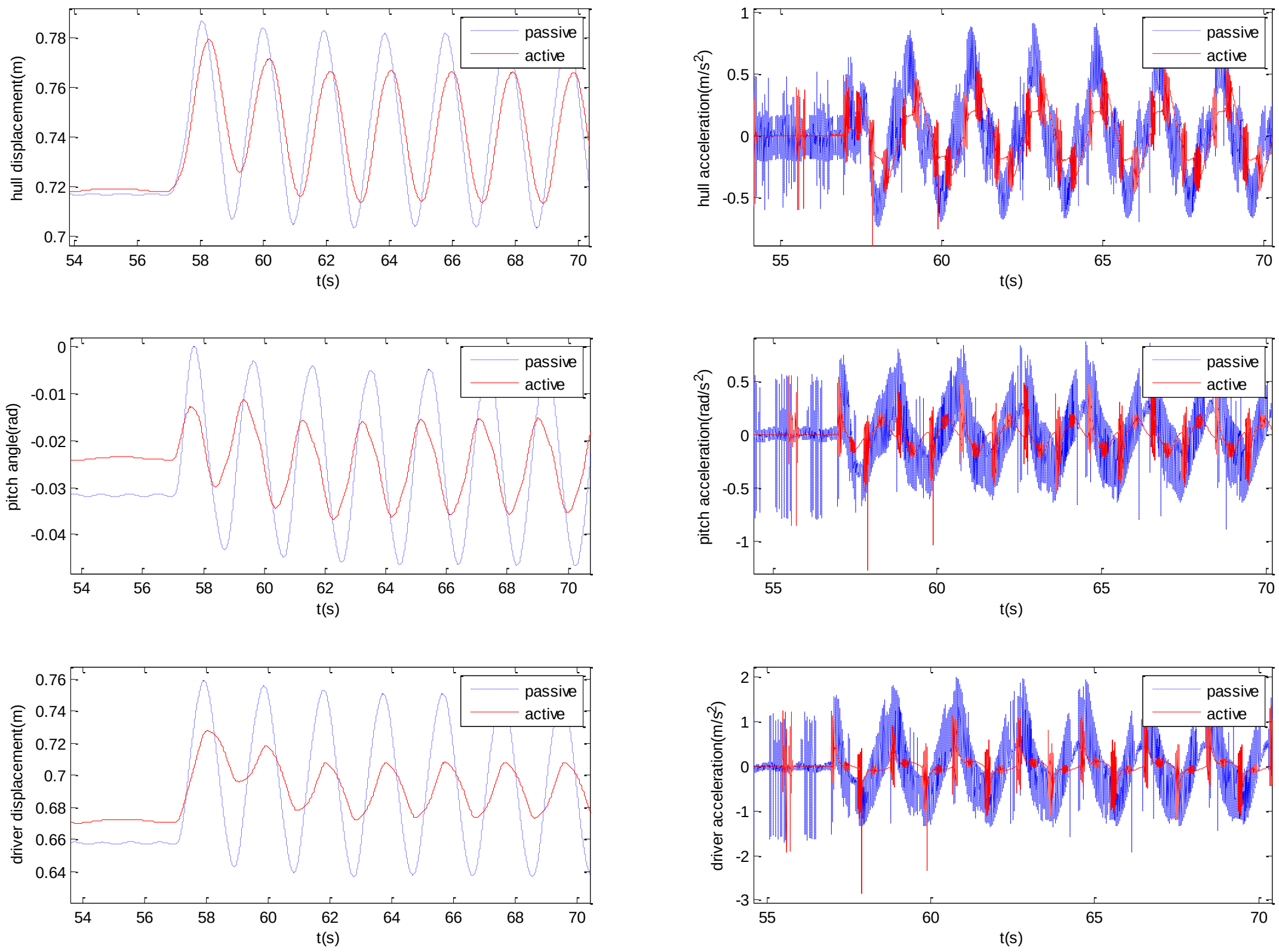

Figure 21. Time histories for the vehicle body heave, pitch and driver's vertical motion and corresponding accelerations for the sinusoidal road input (velocity = 25km/h)

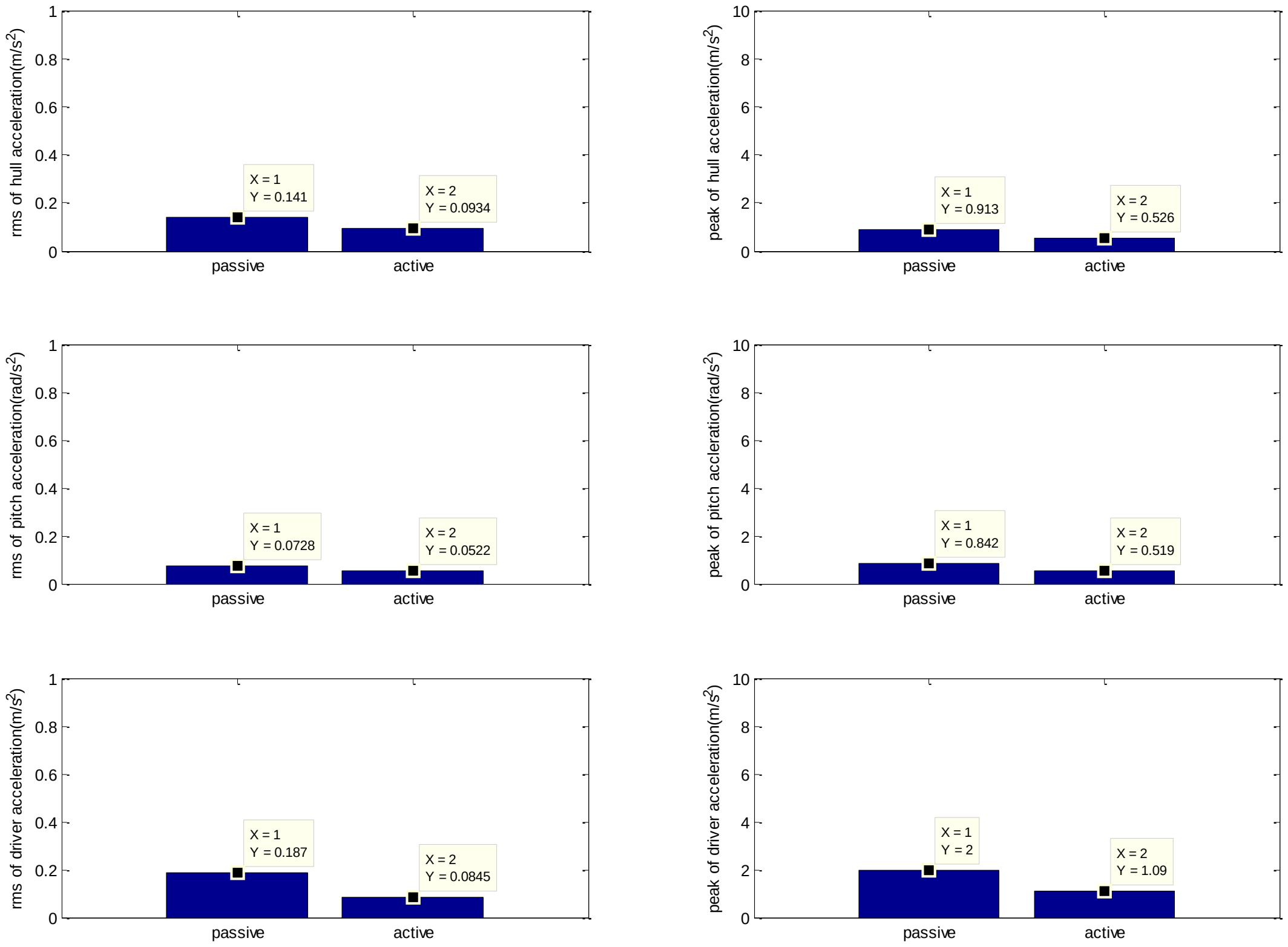

Figure 22. The RMS and peak values for the heave, pitch and driver accelerations for the sinusoidal road input (velocity = 25km/h)

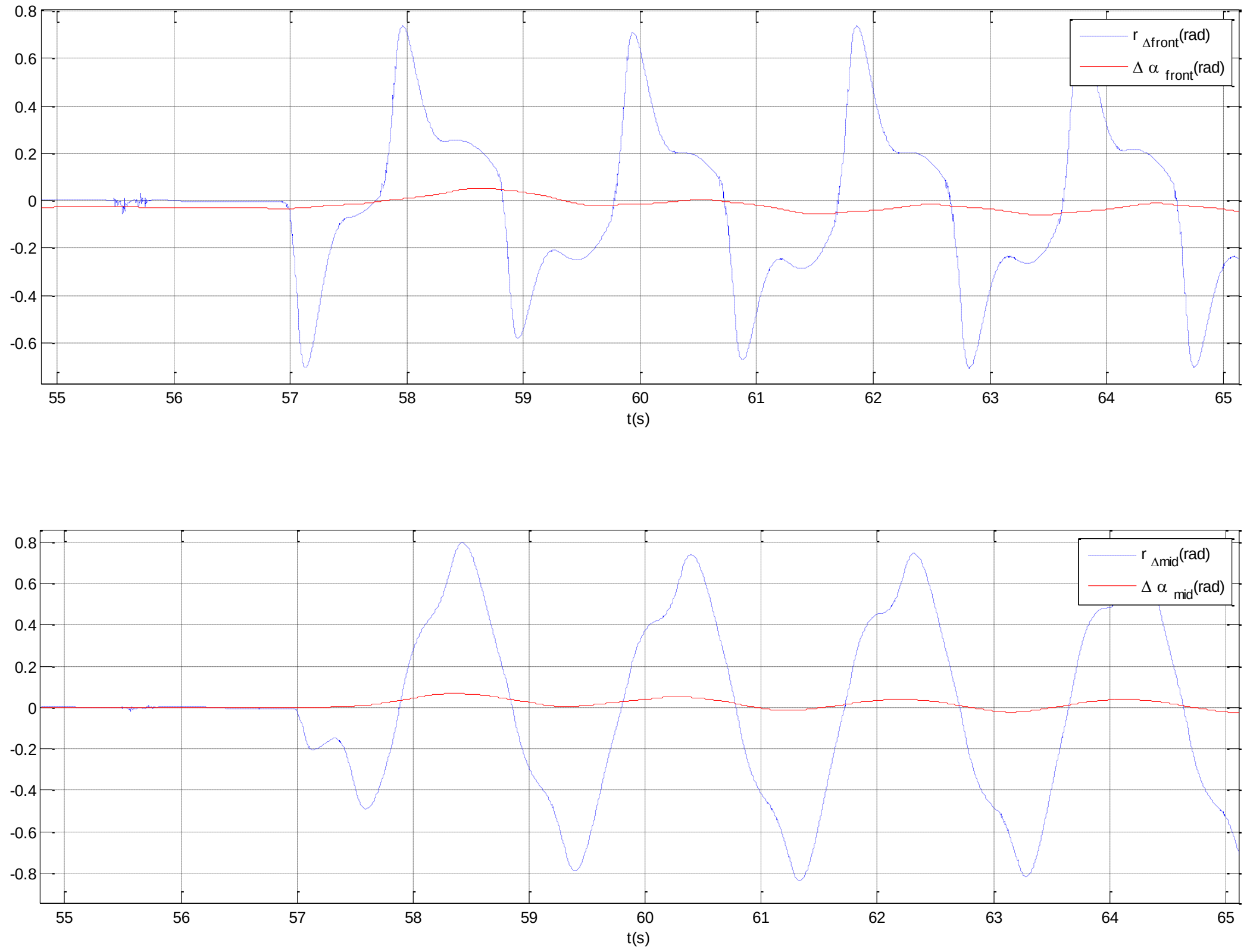

Figure 23. Reference trajectory and corresponding actual arm angle on sinusoidal road input (velocity = 25km/h)

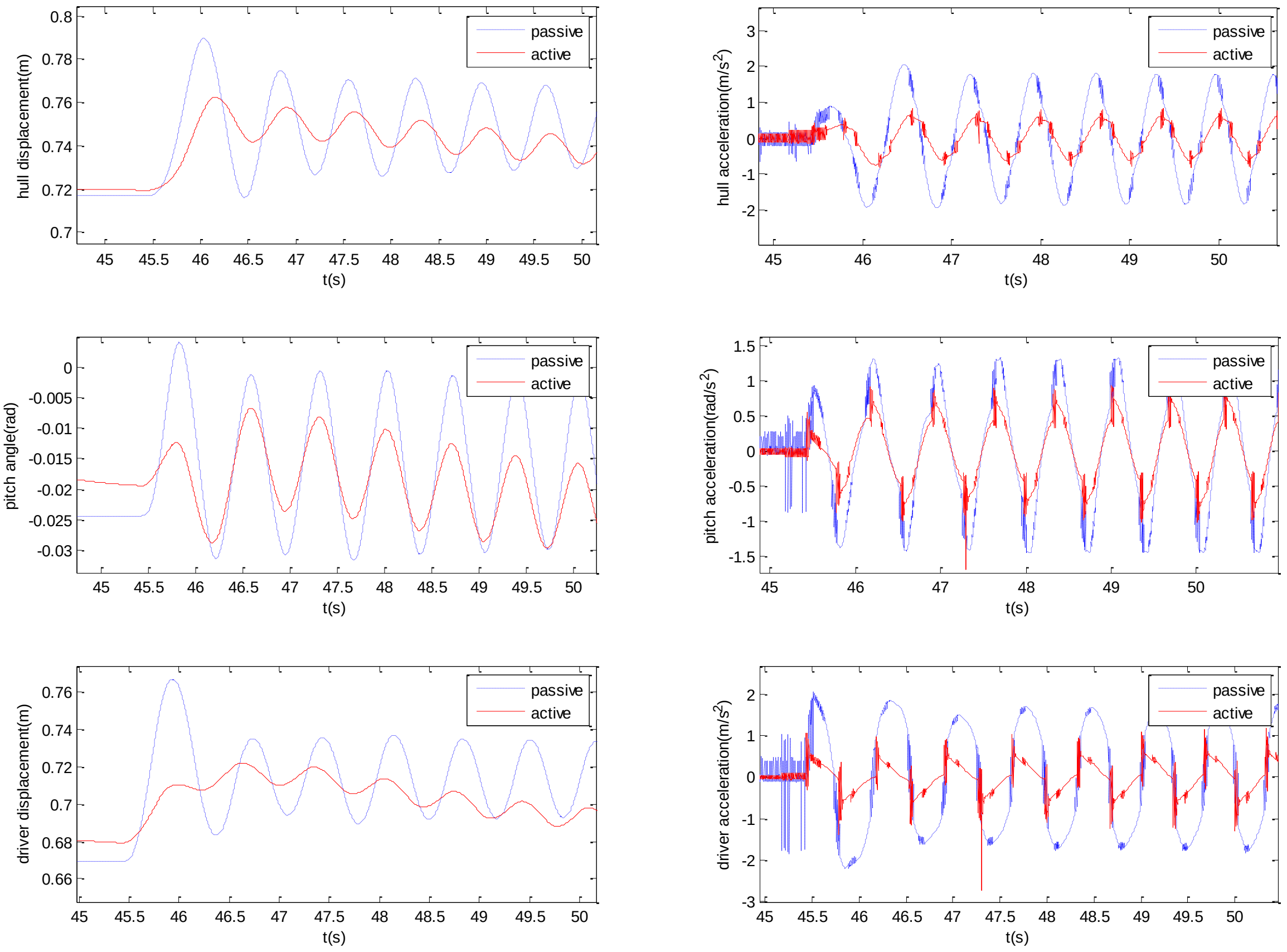

Figure 24. Time histories for the vehicle body heave, pitch and driver's vertical motion and corresponding accelerations for the sinusoidal road input (velocity = 70km/h)

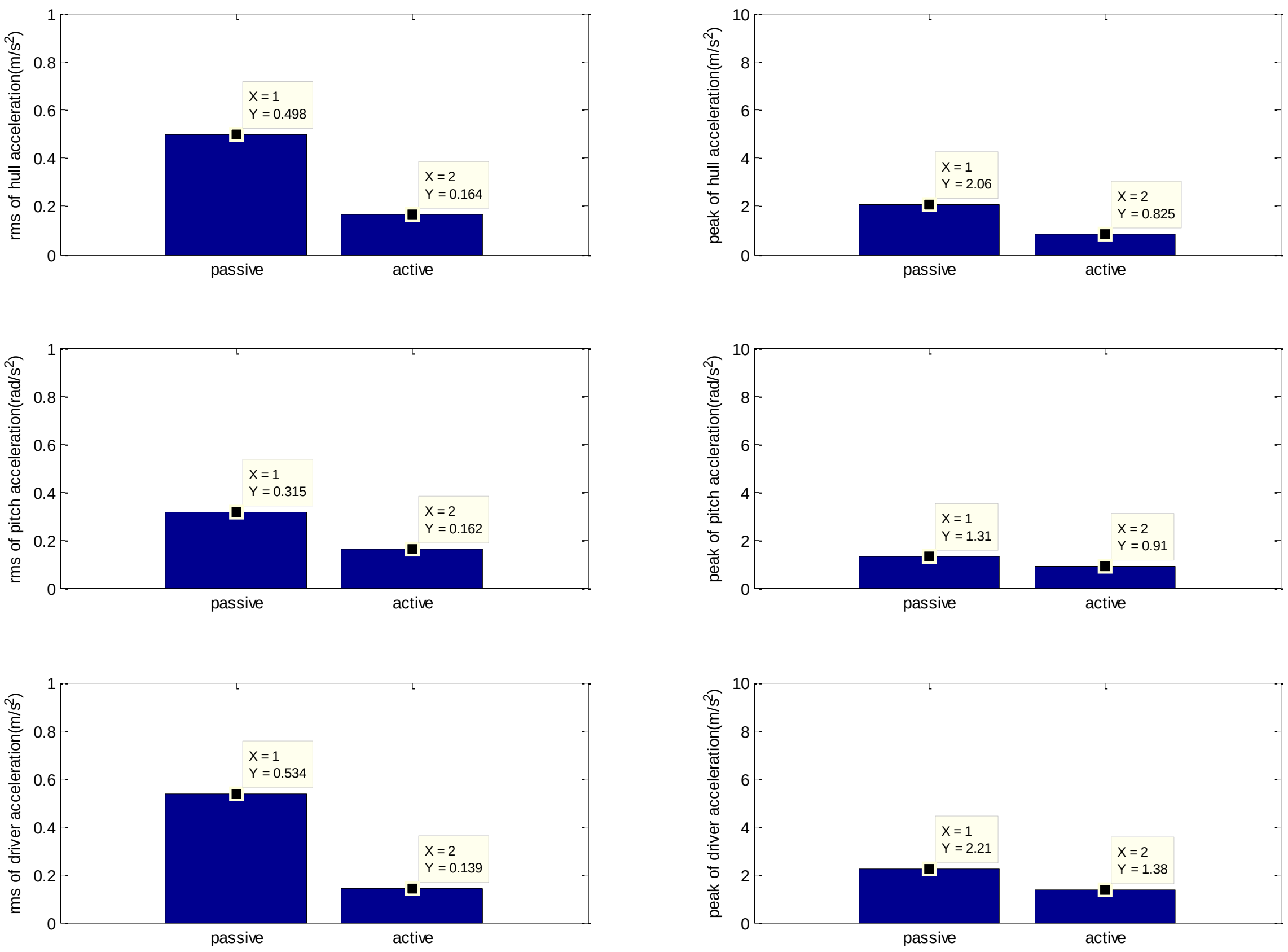

Figure 25. The RMS and peak values for the heave, pitch and driver accelerations for the sinusoidal road input (velocity = 70km/h)

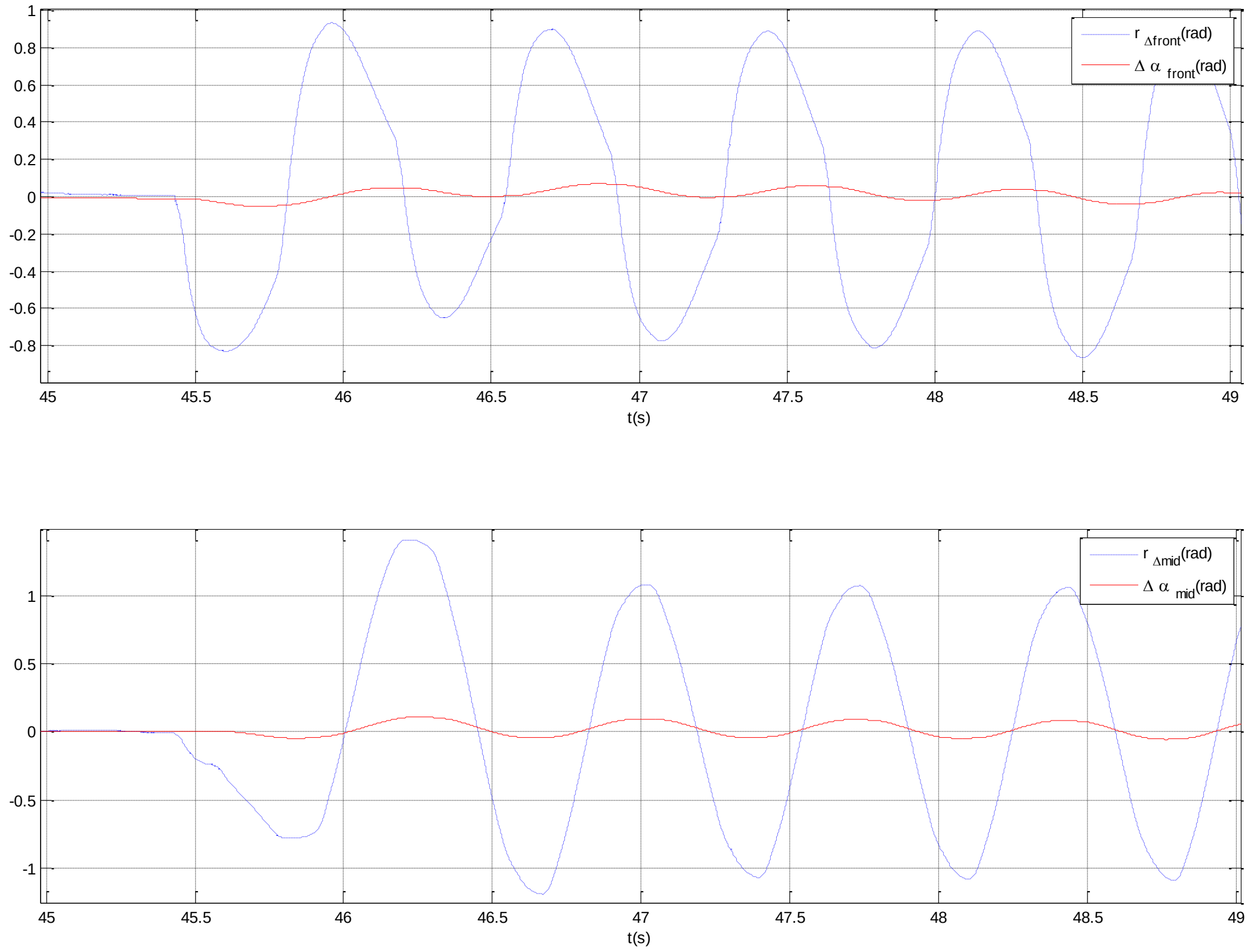

Figure 26. Reference trajectory and corresponding actual arm angle on sinusoidal road input (velocity = 70km/h)

Time responses for the driver and crew member's vertical displacements and corresponding accelerations for the sinusoidal road input are given in Figure 21. The improvement in the ride quality with the active suspension increases substantially as the velocity of the vehicle increases, as seen in Figure 24 and Figure 25. Substantial decreases in frequency weighted RMS values demonstrate the usefulness of the suggested control method.

Figure 21 shows the weighted RMS values of the heave, pitch and driver acceleration of the vehicle body. The frequency weighted RMS values have decreased between 28.36 % and 54.87 % for the vehicle with speed of 25km/h, and 48.57% and 73.97% for the vehicle with speed of 70km/h. Both results for different vehicle speeds prove the stability of the suggested algorithm. Figure 23 shows the reference trajectory and corresponding actual arm angles in the speed of 25km/h, while Figure 26 shows it in the speed of 70km/h.

## V. CONCLUSION

In this paper, first a non-linear quarter car vehicle model was derived and linearized around the equilibrium point. Then it used adaptive law to estimate the plant parameter continuously, since there exist some correlation between each wheel that should have taken into account. In particular, we omitted using heave displacement as the measurable state, since not only it is costly, but also its inaccuracy. According to the simulation results, the ameliorated adaptive linear quadratic control methods gave better ride quality performance. Improvement in ride quality at passenger's position is pale in comparison with that in driver's position since the behavior of last wheel, which is not controlled with additional actuator but affects greatly in passenger's ride quality, is not improved. Extensive responses with different road conditions have indicated that the suggested ALQ controller, without using heave sensor, is likely to provide fairly higher shooting accuracy/operational efficiency and an increased rate of survivability in combat situations.

## REFERENCES

[1] Howard, Geoffrey, John Peter Whitehead, and Donald Bastow. "Car Suspension and Handling." *Warrendale, PA: Society of Automotive Engineers, 2004. 474*(2004).

[2] Rakheja, S., M. F. R. Afonso, and S. Sankar. "Dynamic analysis of tracked vehicles with trailing arm suspension and assessment of ride vibrations."*International Journal of Vehicle Design* 13.1 (1992).

[3] Rae, W. J., and E. M. Kasprzak. "Kinematics of a double A-arm suspension, using Euler orientation variables." *SAE 2002 World Congress Detroit, Michigan*. 2002.

[4] Peicheng, Shi, et al. "Kinematic analysis of a McPherson independent suspension." *Journal of Mechanical Transmission* 1 (2008): 028.

[5] Deo, Hrishikesh V., and Nam P. Suh. "Axiomatic Design of Automobile Suspension and Steering Systems: Proposal for a novel six-bar suspension."*SAE SP* (2004): 189-198.

[6] Jazar, Reza N. *Vehicle dynamics: theory and application*. Springer, 2008.

[7] Cronje, Paul Hendrik, and Pieter Schalk Els. "Improving off-road vehicle handling using an active anti-roll bar." *Journal of Terramechanics* 47.3 (2010): 179-189.

[8] Rajamani, R. "Vehicle dynamics and control, 2006." 387-432.

[9] Hrovat, Davor. "Survey of advanced suspension developments and related optimal control applications." *Automatica* 33.10 (1997): 1781-1817.

[10] Kim, Kiduck, and Doyoung Jeon. "Vibration suppression in an MR fluid damper suspension system." *Journal of intelligent material systems and structures*10.10 (1999): 779-786.

[11] Petek, NiCholas K., et al. "Demonstration of an automotive semi-active suspension using electrorheological fluid." *SAE PUBLICATION SP 1074. NEW DEVELOPMENTS IN VEHICLE DYNAMICS, SIMULATION, AND SUSPENSION SYSTEMS (SAE TECHNICAL PAPER 950586)* (1995).

[12] Roh, H-S., and PARK YOUNGJIN. "Preview control of active vehicle suspensions based on a state and input estimator." *SAE transactions* 107.6 (1998): 1713-1720.

[13] Araki, Yoshiaki, Masahiro Oya, and Hiroshi Harada. "Preview Control of Active Suspension Using Disturbance of Front Wheel." *International Symposium on Advanced Vehicle Control (1994: Tsukuba-shi, Japan). Proceedings of the International Symposium on Advanced Vehicle Control 1994*. 1994

[14] LIAN, R. Enhanced Adaptive Self-Organizing Fuzzy Sliding-Mode Controller for Active Suspension Systems. 2013.

[15] RYU, Seongpil; PARK, Youngjin; SUH, Moonsuk. Ride quality analysis of a tracked vehicle suspension with a preview control. *Journal of Terramechanics*, 2011, 48.6: 409-417.

[16] WONG, J. Y. Dynamics of tracked vehicles. *Vehicle system dynamics,* 1997, 28.2-3: 197-219.

[17] Edara, Ramesh B. "Effective use of multibody dynamics simulation in vehicle suspension system development." (2004).

[18] Ioannou, Petros A., and Jing Sun. *Robust adaptive control*. DoverPublications. com, 2012.

[19] RUSSELL, Stuart. *Artificial intelligence: A modern approach, 2/E*. Pearson Education India, 2003.

[20] CERONE, Vito. *Linear quadratic control: An introduction*. Prentice Hall, 1995.

[21] OLVER, Peter J.; SHAKIBAN, Chehrzad. *Applied linear algebra*. Prentice Hall, 2006.

[22] STRANG, Gilbert. Linear Algebra and Its Applications. Thomson–Brooks. *Cole, Belmont, CA, USA*, 2005.

[23] SUN, Weichao, et al. Adaptive Backstepping Control for Active Suspension Systems With Hard Constraints. 2013.

[24] International Organization for Standardization. *Mechanical vibration and shock-Evaluation of human exposure to whole-body vibration-Part 1: General requirements*. The Organization, 1997.

[25] Paddan, G. S., and M. J. Griffin. "Evaluation of whole-body vibration in vehicles." *Journal of Sound and Vibration* 253.1 (2002): 195-213.

[26] YAGIZ, Nurkan; HACIOGLU, Yuksel. Backstepping control of a vehicle with active suspensions. *Control Engineering Practice*, 2008, 16.12: 1457-1467

[27] Yagiz, Nurkan, Ismail Yuksek, and Selim Sivrioglu. "Robust control of active suspensions for a full vehicle model using sliding mode control." *JSME international journal. Series C, Mechanical systems, machine elements and manufacturing* 43.2 (2000): 253-258..

[28] ZHANG, Xinjie, et al. Semi-Active Suspension Adaptive Control Strategy Based on Hybrid Control. In: *Proceedings of the FISITA 2012 World Automotive Congress*. Springer Berlin Heidelberg, 2013. p. 625-632.

[29] LIN, Jung-Shan; HUANG, Chiou-Jye. Nonlinear backstepping active suspension design applied to a half-car model. *Vehicle system dynamics*, 2004, 42.6: 373-393.